\documentclass[twocolumn]{aastex61}

\usepackage{graphicx}
\usepackage{epstopdf}
\usepackage{comment}
\usepackage{color}
\usepackage{amsmath}

\usepackage[]{hyperref}

\hypersetup{%
  colorlinks=true,
  linkcolor=red,
  citecolor=blue,
  bookmarksopen=true,
  bookmarksnumbered=true,
  pdfstartpage={1},  
  pdftitle = {},
  pdfsubject = {},
  pdfauthor = {} ,
  pdfkeywords = {},
  pdfproducer = {LaTeX with hyperref}
}

\def\Bmax{\ifmmode{\>\vert B\vert_{\text{max}}}\else{$\vert B\vert_{\text{max}}$}\fi}
\def\Bnine{\ifmmode{\>\vert B\vert_{90}}\else{$\vert B\vert_{90}$}\fi}

\def\nH{\ifmmode{\>n_{\textnormal{\sc h}}} \else{$n_{\textnormal{\sc h}}$}\fi}

\def\mG{\ifmmode{\>\mu\mathrm{G}}\else{$\mu$G}\fi}
\def\erg{\ifmmode{\> {\rm erg}}\else{erg}\fi}
\def\keV{\ifmmode{\> {\rm keV}}\else{keV}\fi}

\def\deg{\ifmmode{\>^{\circ}}\else{$^{\circ}$}\fi}
\def\onedeg{\ifmmode{\>1^{\circ}}\else{$1^{\circ}$}\fi}

\def\xvir{\ifmmode{\>x_{vir}}\else{$x_{vir}$}\fi}
\def\Mvir{\ifmmode{\>M_{vir}}\else{$M_{vir} $}\fi}
\def\rvir{\ifmmode{\>r_{vir}}\else{$r_{vir}$}\fi}
\def\vvir{\ifmmode{\>v_{vir}}\else{$v_{vir}$}\fi}

\def\tratio{\ifmmode{\>\tau}\else{$\tau$}\fi}

\def\rms{\ifmmode{\>r_{\textnormal{\sc ms}}}\else{$r_{\textnormal{\sc ms}}$}\fi}

\def\Mpc{\ifmmode{\>{\rm Mpc}} \else{Mpc}\fi}
\def\kpc{\ifmmode{\>{\rm kpc}} \else{kpc}\fi}
\def\pc{\ifmmode{\>{\rm pc}} \else{pc}\fi}

\def\Gyr{\ifmmode{\>{\rm Gyr}} \else{Gyr}\fi}
\def\Myr{\ifmmode{\>{\rm Myr}} \else{Myr}\fi}
\def\yr{\ifmmode{\>{\rm yr}} \else{yr}\fi}
\def\pyr{\ifmmode{\>{\rm yr}^{-1}}\else{yr $^{-1}$} \fi}
\def\s{\ifmmode{\>{\rm s}}\else{s}\fi}
\def\ps{\ifmmode{\>{\rm s}^{-1}}\else{s$^{-1}$}\fi}
\def\Hz{\ifmmode{\>{\rm Hz}}\else{Hz}\fi}

\def\kms{\ifmmode{\>{\rm km\,s}^{-1}}\else{km~s$^{-1}$}\fi}

\def\K{\ifmmode{\>{\rm K}}\else{K}\fi}

\def\sr{\ifmmode{\>{\rm sr}}\else{sr}\fi}
\def\psr{\ifmmode{\>{\rm sr}^{-1}}\else{sr$^{-1}$}\fi}
\def\arcs{\ifmmode{\>{\rm arcsec}}\else{arcsec}\fi}
\def\parcs{\ifmmode{\>{\rm arcsec}^{-1}}\else{arcsec${-1}$}\fi}
\def\parcss{\ifmmode{\>{\rm arcsec}^{-2}}\else{arcsec${-2}$}\fi}

\def\cm{\ifmmode{\>{\rm cm}}\else{cm}\fi}
\def\cc{\ifmmode{\>{\rm cm}^{3}}\else{cm$^{3}$}\fi}
\def\sqc{\ifmmode{\>{\rm cm}^{2}}\else{cm$^{2}$}\fi}
\def\pcc{\ifmmode{\>{\rm cm}^{-3}}\else{cm$^{-3}$}\fi}
\def\psc{\ifmmode{\>{\rm cm}^{-2}}\else{cm$^{-2}$}\fi}

\def\g{\ifmmode{\>{\rm g}}\else{g}\fi}
\def\Msun{\ifmmode{\>{\rm M}_{\odot}}\else{M$_{\odot}$}\fi}
\def\hMsun{\ifmmode{\> h^{-1}{\rm M}_{\odot}}\else{$h^{-1}$M$_{\odot}$}\fi}

\def\Zsun{\ifmmode{\>{\rm Z}_{\odot}}\else{Z$_{\odot}$}\fi}

\def\rayl{\ifmmode{\>{\rm R}}\else{R}\fi}
\def\mR{\ifmmode{\>{\rm mR}}\else{mR}\fi}

\renewcommand{\ion}[2]{\hbox{#1\,{\sc #2}}}

\def\lya{\ifmmode{\>{\rm Ly}\alpha}\else{Ly$\alpha$}\fi}

\def\Ha{\ifmmode{\>{\rm H}\alpha}\else{H$\alpha$}\fi}
\def\Hb{\ifmmode{\>{\rm H}\beta}\else{H$\beta$}\fi}

\def\HI{\ifmmode{\> \textnormal{\ion{H}{i}}} \else{\ion{H}{i}}\fi}
\def\HII{\ifmmode{\> \textnormal{\ion{H}{ii}}} \else{\ion{H}{ii}}\fi}
\def\CIV{\ifmmode{\> \textnormal{\ion{C}{iv}}} \else{\ion{C}{iv}}\fi}
\def\SiIV{\ifmmode{\> \textnormal{\ion{S}{iv}}} \else{\ion{Si}{iv}}\fi}

\def\NHI{\ifmmode{\> {\rm N}_{\HI}} \else{N$_{\HI}$}\fi}
\def\MHI{\ifmmode{\> {\rm M}_{ \HI}} \else{M$_{\HI}$}\fi}

\def\mua{\ifmmode{\>\mu_{ \textnormal{\Ha}}}\else{$\mu_{ \textnormal{\Ha}}$}\fi}
\def\alphabha{\ifmmode{\>\alpha_{B}^{(\textnormal{\Ha})}}\else{$\alpha_{B}^{(\textnormal{\Ha})}$}\fi}

\begin{document}
\defcitealias{marinacci10}{M10}
\defcitealias{gritton17}{GSG17}

\title{Magnetic fields in the Galactic halo restrict fountain-driven recycling and accretion}

\author{Asger Gr\o nnow}
\affiliation{Sydney Institute for Astronomy, School of Physics A28, The University of Sydney, NSW 2006, Australia}

\author{Thor Tepper-Garc\'{\i}a}
\affiliation{Sydney Institute for Astronomy, School of Physics A28, The University of Sydney, NSW 2006, Australia}

\author{Joss Bland-Hawthorn}
\affiliation{Sydney Institute for Astronomy, School of Physics A28, The University of Sydney, NSW 2006, Australia}
\affiliation{ARC Centre of Excellence for Astronomy in Three Dimensions (ASTRO-3D), Sydney, NSW, Australia}

\received{2018 May 10}
\revised{2018 August 7}
\accepted{2018 August 11}

\correspondingauthor{Asger Gr\o nnow}
\email{asger.gronnow@sydney.edu.au}

%--------------------------------------------------------------------------------------------------------------------------------------------------------------------------------
\begin{abstract}
The Galactic halo contains a complex ecosystem of multiphase intermediate-velocity and high-velocity gas clouds whose origin has defied clear explanation. They are generally believed to be involved in a Galaxy-wide recycling process, either through an accretion flow or a large-scale fountain flow, or both.
We examine the evolution of these clouds in light of recent claims that they may trigger condensation of gas from the Galactic corona as they move through it. We measure condensation along a cloud's wake, with and without the presence of an ambient magnetic field, using two- (2D) and three-dimensional (3D), high-resolution simulations. We find that 3D simulations are essential to correctly capture the condensation in all cases. Magnetic fields significantly inhibit condensation in the wake of clouds at $t \gtrsim 25$ Myr, preventing the sharp upturn in cold gas mass seen in previous non-magnetic studies. The magnetic field suppresses the Kelvin-Helmholtz instability responsible for the ablation and consequent mixing of a cloud with halo gas which drives the condensation. This effect is universal across different cloud properties (density, metallicity, velocity) and magnetic field properties (strength and orientation). Simple convergence tests demonstrate that resolving the gas on progressively smaller scales leads to even less condensation. While condensation still occurs in all cases, our results show that an ambient magnetic field drastically lowers the efficiency of fountain-driven accretion and likely also accretion from condensation around high-velocity clouds. These lower specific accretion rates are in better agreement with observational constraints compared to 3D, non-magnetic simulations.
\end{abstract}

%--------------------------------------------------------------------------------------------------------------------------------------------------------------------------------

%%%%%%%%%%%%%%%%%%%%%%%%%%%%%%%%%%%%%%%%%%%%%%%%%%%%%%%%%%%%%%%%%%%%%%
%%%%%%%%%%%%%%%%%%%%%%%%%%%%%%%%%%%%%%%%%%%%%%%%%%%%%%%%%%%%%%%%%%%%%%

%--------------------------------------------------------------------------------------------------------------------------------------------------------------------------------
\section{Introduction}
\label{sec:intro}
The Galaxy is surrounded by large amounts of neutral hydrogen (\textsc{Hi}) gas structures, collectively referred to as \textsc{Hi} High-velocity Clouds (HVCs; \citealt{muller63}). These include the Magellanic Stream \citep[MS; ][]{mathewson74}, a tail of gas that extends for more than $200^\circ$ on the sky from the Magellanic Clouds \citep{nidever10}. HVCs together are estimated to contain a neutral gas mass of $\sim 10^7 ~M_\odot$ \citep{putman12}. Their total gas mass may in fact be a factor of 2 higher \citep{lehner12}. The MS alone contains a total mass of order $\sim 10^9 (d / 55 ~{\rm kpc}) ~M_\odot$, depending on its distance $d$ \citep{fox04,bland-hawthorn07}. This gas may be enough to sustain the average Galactic star formation (SF) at the current observed rates of $\sim 1 ~M_\odot$ yr$^{-1}$ \citep{robitaille10}, if accreted within a few Gyr.

An alternative source of gas is the Galactic corona. Indeed, a number of observations support the existence of an extended ($r \gtrsim$ 50 kpc), diffuse ($n \sim 10^{-5} - 10^{-3}$ cm$^{-3}$), massive ($\sim 10^{11} ~M_\odot$) \citep{faerman17} halo of hot ($T \sim 10^6$ K) gas around the Galaxy: (i) the mere existence of HVCs \citep[][]{spitzer56}; (ii) their observed head-tail morphology \citep{putman11}; (iii) the extended X-ray emission around the Galaxy \citep{li17}. Although the precise extent, structure, and origin of the hot halo are unknown, it is generally believed that it represents a substantial reservoir of baryons \citep{bregman07}.

It has been suggested that gas could be forced out from the corona and transported onto the disk, thus allowing for Galactic SF to be sustained over significant periods. However, the mechanism by which this accretion occurs is not entirely clear. One possibility invokes the motion of H\textsc{i} Intermediate-velocity Clouds (IVCs) as the relevant trigger\textemdash that is, gas cloudlets ejected by supernova explosions from the disk in a processes dubbed the `Galactic fountain' \cite[][]{shapiro76}, which reach heights of up to a few kpc above the Galactic plane before falling back to the disk.

Early work on fountain-triggered gas condensation \citep[][hereafter \citetalias{marinacci10}; see also \citealt{armillotta16} and references therein]{marinacci10} suggests that the gas ablated from IVCs while they move through the disk-halo interface and across the lower halo effectively seeds condensation of the ambient gas as the ablated material mixes with the ambient medium. Indeed, as the hot, metal-poor halo gas mixes with the denser, higher metallicity gas ablated from IVCs, its cooling time is greatly reduced.\footnote{The gas cooling rate scales with the gas metallicity ($Z$) and the gas density ($n$) as $\sim Z n^2$. The cooling timescales in turn scales with $\Lambda$, $n$, and the gas temperature ($T$) as $\sim n ~T / \Lambda \sim T / n Z $. } While the gas cools, it becomes denser, thus cooling further until it can no longer be supported by the ambient gas pressure and falls onto the disk. For the gas to actually rain down onto the plane, the density contrast of the cloudlets is required to be large enough for nonlinear effects to become important \citep{binney09,joung12}. \textsc{Hi} disks in spiral galaxies are likely to extend twice as far as currently observed \citep{bland-hawthorn17}. The clumping of this outer gas aids its survival against photoionization from the cosmic ultraviolet background radiation. These outer gas disks show that gas settles gently to the disk presumably along a flow roughly parallel with the disk. The SF rates are very low at these large radii, and so the fountain process is mainly associated with the inner disk.

The idea of halo gas condensation seeded by traversing clouds is not restricted to IVCs, however. \citet[hereafter \citetalias{gritton17}]{gritton17} investigated whether the more distant HVCs could also lead to condensation of halo material in their wakes. They found that HVCs were more efficient than IVCs at triggering condensation of the ambient gas due to their greater mass and contact area \citep[i.e. an increased turbulent mixing layer surface;][]{begelman90}.

However significant, all previously discussed work focusing on gas condensation has systematically ignored the effect of the magnetic field. A magnetic field is known to exist throughout the disk and expected to be present in the halo as well \citep{pshirkov11,beck15}. Crucially, magnetic fields have been shown to slow down the effect of mixing by dampening the Kelvin-Helmholtz (KH) instability that drives the turbulent mixing \citep{sur14}. This result has been confirmed by numerical studies of cloud-wind interaction \citep[e.g.][]{jones96,mccourt15,banda-barragan16,goldsmith16,gronnow17}. Also, magnetic fields appear to have a positive impact on the development of thermal instabilities leading to fragmentation of an uniform medium into a multiphase medium \citep{ji18}. Therefore, it is reasonable to suspect that magnetic fields will affect the condensation of gas seeded by moving clouds. This effect should be particularly relevant in the case of the clouds moving through the lower halo where the magnetic field is relatively strong \citep[$\sim 5 ~\mu{\rm G}$,][]{jansson12a}.

%--------------------------------------------------------------------------------------------------------------------------------------------------------------------------------
\begin{deluxetable*}{lccccccc}[htb]
\tablewidth{\textwidth} 

\tablecaption{Fixed simulation parameters \label{table:ics}}

\tablehead{\colhead{$v_{\text{wind}}$\tablenotemark{$a$}} & \colhead{$n_h$} & \colhead{$T_h$} & \colhead{[Fe/H]$_{\text{h}}$} & \colhead{$r_c$} & \colhead{$x$} & \colhead{$y$}  & \colhead{$z$}\\ 
\colhead{(\kms)} & \colhead{( \pcc)} & \colhead{(K)} & \colhead{} & \colhead{(kpc)} & \colhead{(kpc)} & \colhead{(kpc)} & \colhead{(kpc)} }

\startdata
75 & $10^{-3}$ & $2 \times 10^6$ & -0.5 & 0.1 & $-2.0 \leq x \leq 10.0$ & $-0.6 \leq y \leq 0.6$ & $-0.6 \leq z \leq 0.6$ \\
\enddata
\tablenotetext{a}{This corresponds to a Mach number of 0.45.}
\end{deluxetable*}
%--------------------------------------------------------------------------------------------------------------------------------------------------------------------------------

The goal of this paper is to investigate in detail the effect of the Galactic magnetic field on the condensation of ambient gas along the wakes of intermediate-velocity (and a few high-velocity) fountain gas clouds through high-resolution 3D magnetohydrodynamic (MHD) simulations. In essence, we find that the presence of a magnetic field significantly reduces the amount of condensation compared to earlier work based on pure HD simulations. Also, we argue that 2D simulations are not adequate because they artificially lower the amount of condensation. Thus, full 3D simulations are necessary to correctly capture the condensation process. We show that the loss of this artificial suppression of condensation in our 3D simulations is compensated by the physical suppression by the magnetic field and that accretion rates in our magnetic simulations are in good agreement with observational constraints.

The paper is organized as follows. In Section~\ref{sect:ICs} we describe the numerical setup. In Section~\ref{sec:results} we show visualizations of the simulations and describe the efficiency of condensation in our simulations with and without magnetic fields and various densities and metallicities. In Section~\ref{sec:discussion} we discuss the wider implications for accretion through condensation in the wake of clouds with comparisons to previous studies as well as limitations. Finally, we conclude our findings in Section~\ref{sec:summary}.

%%%%%%%%%%%%%%%%%%%%%%%%%%%%%%%%%%%%%%%%%%%%%%%%%%%%%%%%%%%%%%%%%%%%%%
%%%%%%%%%%%%%%%%%%%%%%%%%%%%%%%%%%%%%%%%%%%%%%%%%%%%%%%%%%%%%%%%%%%%%%

%--------------------------------------------------------------------------------------------------------------------------------------------------------------------------------
\section{Numerical experiments}
\label{sect:ICs}

%--------------------------------------------------------------------------------------------------------------------------------------------------------------------------------
\begin{deluxetable*}{lcccccl}
\tabletypesize{\scriptsize}
\tablewidth{\textwidth}
\tablecaption{Varying Simulation Parameters. \label{table:ics2}}
\tablecolumns{7}

\tablehead{
\colhead{Name\tablenotemark{$a$}} & \colhead{$n_c$\tablenotemark{$b$}} & \colhead{$T_c$\tablenotemark{$c$}} & \colhead{[Fe/H]$_{\text{c}}$} & \colhead{$\vert \mathbf{B}_0\vert$\tablenotemark{$d$}} & \colhead{Maximum resolution\tablenotemark{$e$}} & \colhead{Notes}\\
\colhead{} & \colhead{(\pcc)} & \colhead{(K)} & \colhead{} & \colhead{($\mG$)} & \colhead{(cells/$r_c$)} & \colhead{} }

\startdata
2LS & 0.2 & $10^4$ & -0.5 & 1 & 64 & ...\\
2LW & 0.2 & $10^4$ & -0.5 & 0.1 & 64 & ...\\
2LN & 0.2 & $10^4$ & -0.5 & 0 & 64 & ...\\
2HS & 0.2 & $10^4$ & 0 & 1 & 64 & ...\\
2HW & 0.2 & $10^4$ & 0 & 0.1 & 64 & ...\\
2HN & 0.2 & $10^4$ & 0 & 0 & 64 & ...\\
4LS & 0.4 & $5 \times 10^3$ & -0.5 & 1 & 64 & ...\\
4LW & 0.4 & $5 \times 10^3$ & -0.5 & 0.1 & 64 & ...\\
4LN & 0.4 & $5 \times 10^3$ & -0.5 & 0 & 64 & ...\\
4HS & 0.4 & $5 \times 10^3$ & 0 & 1 & 64 & ...\\
4HW & 0.4 & $5 \times 10^3$ & 0 & 0.1 & 64 & ...\\
4HN & 0.4 & $5 \times 10^3$ & 0 & 0 & 64 & ...\\
\tableline
2HS-h & 0.2 & $10^4$ & 0 & 1 & 128 & High resolution\\
2HN-h & 0.2 & $10^4$ & 0 & 0 & 128 & idem\\
2HS-l & 0.2 & $10^4$ & 0 & 1 & 32 & Low resolution\\
2HN-l & 0.2 & $10^4$ & 0 & 0 & 32 & idem\\
2HS-s & 0.2 & $10^4$ & 0 & 1 & 64 & Static grid\\
2HN-s & 0.2 & $10^4$ & 0 & 0 & 64 & idem\\
2HS-s-ct & 0.2 & $10^4$ & 0 & 1 & 64 & Static grid using constrained transport\\
2HS-ls & 0.2 & $10^4$ & 0 & 1 & 32 & Static grid at low resolution\\
2HN-ls & 0.2 & $10^4$ & 0 & 0 & 32 & idem\\
2HS-ls-ct & 0.2 & $10^4$ & 0 & 1 & 32 & Static grid at low resolution using constrained transport\\
2HS-v200 & 0.2 & $10^4$ & 0 & 1 & 64 & High wind speed ($v_{\text{wind}}=200 \kms$)\\
2HW-v200 & 0.2 & $10^4$ & 0 & 0.1 & 64 & idem\\
2HN-v200 & 0.2 & $10^4$ & 0 & 0 & 64 & idem\\
2HS-45 & 0.2 & $10^4$ & 0 & 1 & 64 & Magnetic field at $45^\circ$ angle w.r.t. velocity in the $xy$-plane\\
2HW-45 & 0.2 & $10^4$ & 0 & 0.1 & 64 & idem\\
2HS-225 & 0.2 & $10^4$ & 0 & 1 & 64 & Magnetic field at 22$\stackrel{\hbox{$\scriptstyle{\circ}$}}{\hbox{.}}$\hspace{-0.6pt}5 angle w.r.t. velocity in the $xy$-plane\\
2HW-225 & 0.2 & $10^4$ & 0 & 0.1 & 64 & idem\\
2HS-par & 0.2 & $10^4$ & 0 & 1 & 64 & Magnetic field parallel to the velocity\\
2HW-par & 0.2 & $10^4$ & 0 & 0.1 & 64 & idem\\
2HS-eq & 0.2 & $10^4$ & 0 & 1 & 64 & Magnetic field with equal components in each direction\\
2HW-eq & 0.2 & $10^4$ & 0 & 0.1 & 64 & idem\\
2LHN & 0.2 & $10^4$ & -0.5 & 0 & 64 & Solar metallicity in halo\\
2HHN & 0.2 & $10^4$ & 0 & 0 & 64 & idem\\
2HN-2D & 0.2 & $10^4$ & 0 & 0 & 64 & 2D AMR grid\\
2HN-2D-h & 0.2 & $10^4$ & 0 & 0 & 128 & 2D AMR grid at high resolution\\
\enddata
\tablecomments{Unless stated otherwise, all simulations are full 3D, AMR at our our standard resolution of 64 cells / $r_c$, adopting a wind speed of $v_{\text{wind}}=75 \kms$}
\tablenotetext{a}{The naming convention is as follows: the number indicates the adopted density in tenths of \pcc, for example, 2 meaning $n_c=0.2 \pcc$; the following letter indicates either low (L) or high (H) metallicity; the third letter indicates the strength of the magnetic field either strong (S), weak (W), or no field (N). The low-density simulations below the line have extra letters indicating further variations.}
\tablenotetext{b}{Initial density contrasts are $\chi=n_c/n_h\approx 200$ for low density clouds and $\chi\approx 400$ for high density clouds. Mass density contrasts $\rho_c/\rho_h$ are approximately a factor of two greater because of differences in mean molecular weight.}
\tablenotetext{c}{The cloud temperature is not a free parameter but rather follows from the density and temperature of the halo and the density of the cloud given cloud-halo pressure equilibrium.}
\tablenotetext{d}{The ratio of gas pressure to magnetic pressure $\beta=8\pi P/\vert \mathbf{B}_0\vert^2$ is approximately 7 (700) in the strong (weak) field simulations.}
\tablenotetext{e}{Given as the number of cells on the highest AMR level per cloud radius.}
\end{deluxetable*}
%--------------------------------------------------------------------------------------------------------------------------------------------------------------------------------

We simulate a cloud moving through the halo as follows. We set up a gas cloud with a prescribed density profile within a three-dimensional (3D), rectangular domain filled with a uniform hot medium of density $n_{\text{h}}$. The pressure is initially uniform so that the cloud is in pressure equilibrium with the hot medium. Initially, the cloud is at rest with respect to the domain's coordinate system, while the surrounding halo gas is moving at a constant velocity $v_{\text{wind}}$ (i.e. in the form of a `wind') parallel to the $x$-axis. We use zero-gradient (`outflow') boundary conditions everywhere, except at the $-x$ injection boundary where quantities are held constant at the wind values. The cloud has a smooth density profile described by
\begin{equation} \label{eq:densprofile}
	n(r)=n_h + \frac{1}{2}(n_c - n_h)\left(1 - \tanh{\left[ s \left(\frac{r}{r_c}-1\right)\right]}\right) \, ,
\end{equation}
where $n$ is total particle density, $r_c$ is the cloud radius, and $s$ sets the steepness of the profile; here we adopt $s=10$. This choice implies $n(r_c) \approx n_c/2$ if $n_c \gg n_h$, which is the case in all our simulations. The velocity and metallicity follow sharp top-hat profiles, with boundaries at $r=1.3 ~r_c$ and the radius at which $n(r) = 2 n_h$, respectively. To keep track of cloud material at later times, we tag the cloud with a passive tracer $C$ which is set to 1 for $n(r) < 2 n_h$ and to 0 elsewhere. Note that because the mean molecular weight ($\mu$) varies with temperature ($T$), the steepness of the {\em mass} density profile, $\rho(r)=n(r)\mu(T)$, will generally differ slightly from $n(r)$. Both the cloud and the halo are taken to be monatomic ideal gases with an adiabatic index $\gamma=5/3$. 

Radiative cooling is included based on the collisional ionization equilibrium cooling tables of \cite{sutherland93}.\footnote{Although these may be somewhat outdated, we adopt them for a meaningful comparison with earlier work. Also, we do not expect the choice of cooling curve to have a significant effect on our general results.} Variations in $\mu$ with $T$, and cooling rate ($\Lambda$) with metallicity ($Z$), are taken into account. The relative elemental abundances are assumed to be solar.

Our parameter values are chosen to generally follow the previous studies of \citetalias{marinacci10} and related papers to facilitate comparisons to earlier work. The halo temperature is initially set to $T_h=2 \times 10^6$ \citep{henley15}, and the halo metallicity to [Fe/H]$_{\text{h}} = -0.5$ \citep{miller15}.

Tab.~\ref{table:ics} lists the parameters and their values that are held fixed during a run (except for a few simulations with higher velocity or halo metallicity), as well as the simulation domain limits. The parameters and their initial values that vary across different simulations are listed in Tab.~\ref{table:ics2}. Note that the choices $n_c=0.2 \pcc$ and [Fe/H]$_c=0$ are the fiducial values used whenever other parameters of the simulations are varied.

To make a connection with observations, we choose a set of densities and metallicities typical for the range of values measured across the population of Galactic halo clouds \citep{wakker01,lehner09,putman12,fox16}. The choice of initial density and density profile implies initial masses of $2.3 \times 10^4 M_\odot$ ($4.9 \times 10^4 M_\odot$) for low (high) density clouds, respectively.

The ambient (coronal) magnetic field, in particular around fountain clouds, however, is not well constrained. For a height above the disk of $z=10$ kpc, the Galactic field models of \cite{sunreich10} and \cite{jansson12a} both yield a magnetic field strength of $\vert\mathbf{B}\vert \approx 0.4 \mu$G at around the Solar galactocentric distance of $R=8$ kpc with values increasing/decreasing by orders of magnitude as $R$ decreases/increases. Based on this, we initially set the field to be uniform throughout the simulation domain pointing in the positive $y$ direction with a weak field, highly super-Alfv\'{e}nic case of $\vert \mathbf{B}_0\vert=0.1 \mu$G and a strong field, slightly sub-Alfv\'{e}nic case of $\vert \mathbf{B}_0\vert=1 \mu$G.\footnote{The Alfv\'{e}nic Mach number is  $\mathcal{M}_A=v_{\text{wind}}/v_A$, where $v_A=\vert \mathbf{B}\vert/\sqrt{4\pi\rho}$ is the Alfv\'{e}n speed.}

The system composed of the cloud, the wind, and the magnetic field is evolved by solving the ideal MHD equations with the code {\sc Pluto} \citep[version 4.1 of the code last described by][]{mignone12}\footnote{Our setup files and modifications to the {\sc Pluto} source code are available upon request from the corresponding author}. The ideal MHD approximation is adequate for our simulations following the arguments made in \cite{gronnow17}. The divergence free condition ($\nabla \cdot \mathbf{B}=0$) is approximately enforced through the hyperbolic divergence cleaning method \citep{dedner02}, which generally dampens the divergence by several orders of magnitude relative to the average field strength \citep{hopkins16} and is computationally efficient. We use the dimensionally unsplit Corner Transport Upwind (CTU) integration scheme \citep{colella90,mignone05} and the HLLC Riemann solver \citep{toro94}. Different Riemann solvers generally lead to different evolution, with less diffusive (i.e. more accurate) solvers unfortunately generally also being less numerically stable \citep{martizzi18}. The HLLC method represents a good compromise between the two.

To achieve a grid spacing small enough to properly follow the evolution of mixing layers while keeping the computational time reasonably low, we make use of the Adaptive Mesh Refinement (AMR) technique, provided as part of the {\sc Pluto} package. Cells in the simulation domain are refined (coarsened) whenever the second derivative error norm of the density exceeds (falls short of) an arbitrary threshold of 0.65.\footnote{This value has been found by trial and error to yield the required resolution while keeping the size of the refined grid reasonable.} All our standard resolution simulations use six levels of refinement, starting from a base (coarse) level with a resolution of $\mathcal{R}=2$ cells/$r_c$ which we denote as $\mathcal{R}_2$. This implies a limiting spatial resolution of $\mathcal{R}=2^5 \times \mathcal{R}_2=\mathcal{R}_{64}$ (i.e. 64 cells/$r_c$). This is equivalent to a minimum linear cell length of $\Delta x \approx 1.6$ pc.

In addition to our standard, fully 3D, MHD simulations with weak and strong transverse fields, we consider the following variations. In order to assess the effect of magnetic fields, for most of our MHD simulations, we run identical simulations where the magnetic field is turned off. We also run simulations with uniform magnetic fields pointing in non-transverse directions in order to assess the effect of different field orientations. We run simulations with higher wind velocities, and simulations with higher halo metallicities. In addition, we run a small set of simulations at different resolutions to assess the convergence of our results and with a static grid to assess the accuracy of using adaptive grids. Two of the static grid simulations additionally also use another more exact method of minimizing magnetic field divergence to evaluate our standard divergence cleaning scheme Finally, we run a pair of 2D, pure HD (i.e. non-magnetic) AMR simulations to test whether 2D simulations are a satisfactory approximation to model this particular type of systems, as claimed in earlier work \citep[e.g. \citetalias{marinacci10}; ][]{armillotta16}.

%--------------------------------------------------------------------------------------------------------------------------------------------------------------------------------
\begin{figure*}[htb]
\begin{center}
\includegraphics[width=\textwidth]{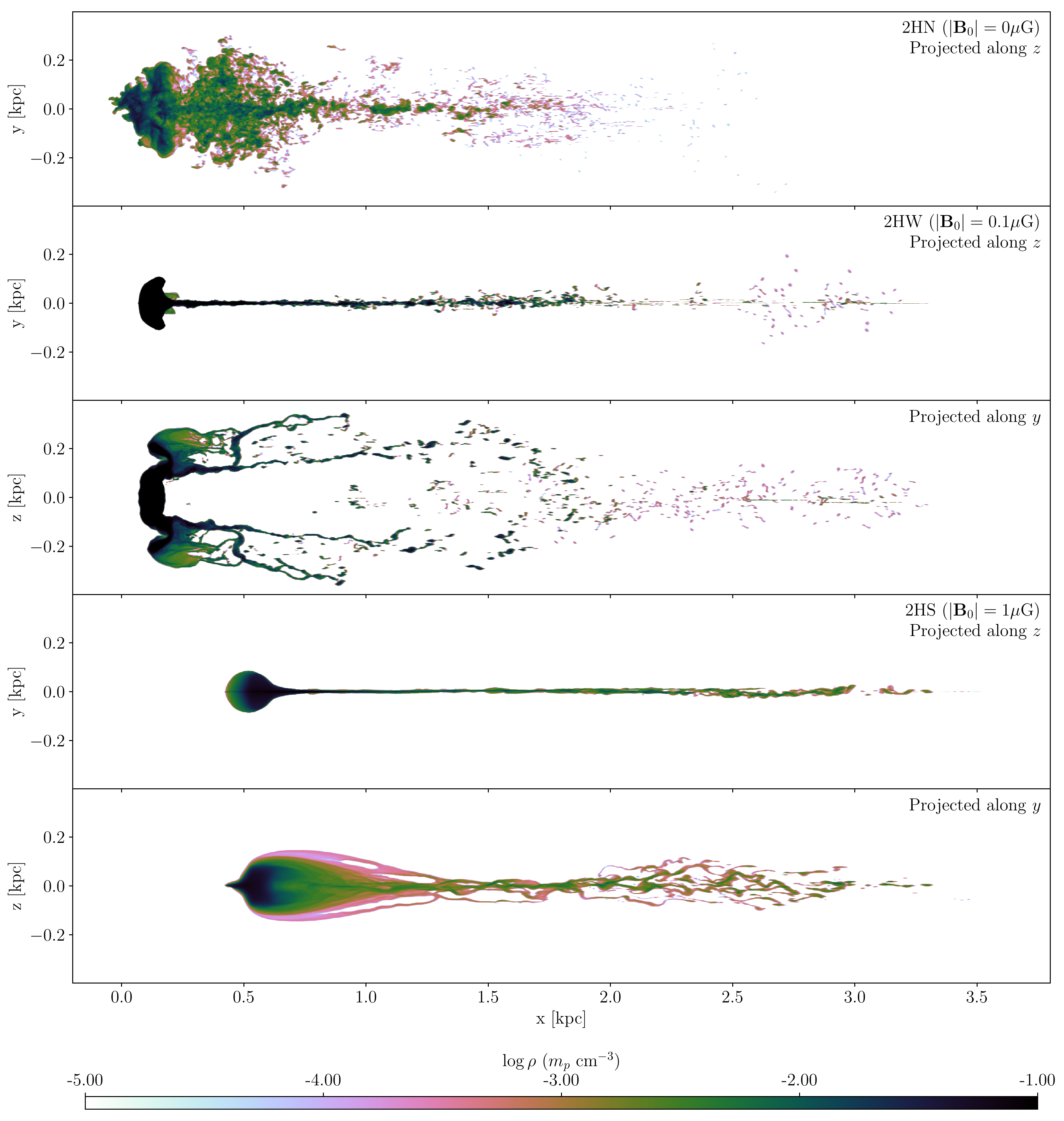}
\end{center}
\caption{Projected mass density of cold ($T < 5 \times 10^5$ K) gas at the end of the simulations at $t\approx 50$ Myr for simulation 2HN (top, projected along the $z$-axis), simulation 2HW (second and third panels from the top, projected along the $z$-axis and the $y$-axis, respectively), and simulation 2HS (fourth and fifth panels from the top, projected along the $z$-axis and the $y$-axis, respectively). Because of the symmetry in the initial conditions, the $y$-projection of the non-magnetic simulation (2HN) is similar to the $z$-projection and is therefore omitted. The cloud in simulation 2HS is at a different position compared to the other two cases because it has been significantly slowed by magnetic tension. The color coding indicates the value of the density in units of $m_p$ cm$^{-3}$ on a logarithmic scale, where $m_p$ is the proton mass.}
\label{fig:rhocoldprojs}
\end{figure*}
%--------------------------------------------------------------------------------------------------------------------------------------------------------------------------------

%--------------------------------------------------------------------------------------------------------------------------------------------------------------------------------
\begin{figure*}[htb]
\begin{center}
\includegraphics[width=\textwidth]{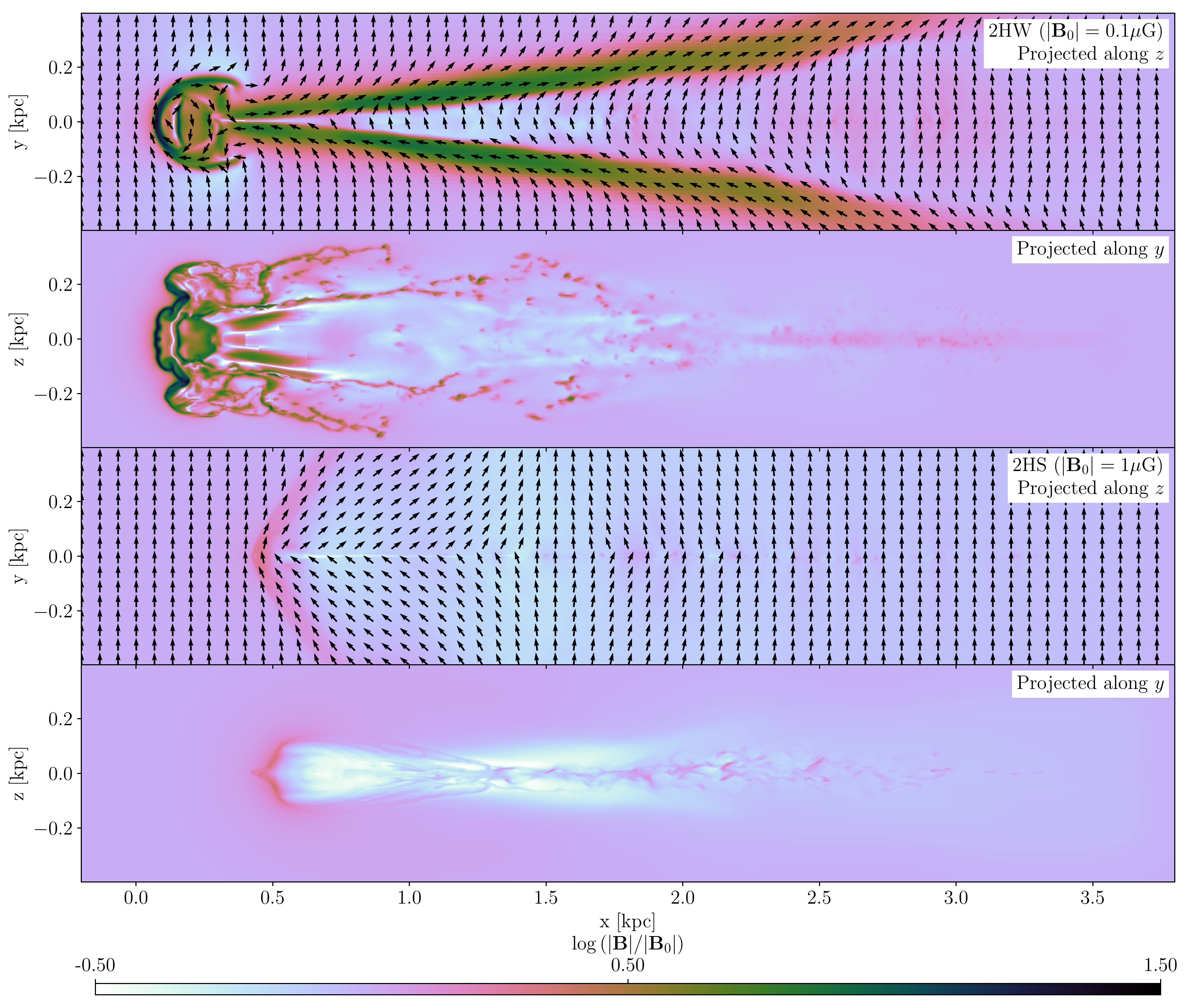}
\end{center}
\caption{Magnetic field strength at the end of the simulations at $t\approx 50$ Myr relative to the initial field strength (i.e. the field amplification) in simulations 2HW (top pair) and 2HS (bottom pair) sliced along $z=0$ (top of each pair) and $y=0$ (bottom of each pair) across the simulation volume. Arrows indicate the direction of the field in the $xy$-plane. In the $xz$-plane, the field points primarily out of the page and arrows are therefore omitted. The color indicates the magnetic field amplification on a logarithmic scale.}
\label{fig:bfieldslices}
\end{figure*}
%--------------------------------------------------------------------------------------------------------------------------------------------------------------------------------

%--------------------------------------------------------------------------------------------------------------------------------------------------------------------------------
\section{Results}
\label{sec:results}

Gas condensation is quantified by measuring the amount of `cold' gas within the simulation domain at any given time, relative to its initial value. We take gas to be `cold' if its temperature $T < 5 \times 10^5$ K, following \citetalias{marinacci10}. This is a relatively high temperature compared to the initial cloud core temperature of $T = 10^4$ K ($5 \times 10^3$ K) for low (high) density clouds. However, none of our results change qualitatively if we instead use a lower threshold such as, for example, $T < 2 \times 10^4$ K as was used in \cite{armillotta16}.

For each run we have verified that the domain in all our simulations is sufficiently large such that no cold gas leaves the volume over the course of the simulation, thus avoiding a bias in the measured condensation. In addition to the visualizations provided here, animations, including 3D renderings, can be found at \href{http://www.physics.usyd.edu.au/~agro5109/animations/condensation.html}{http://www.physics.usyd.edu.au/$\sim$agro5109/\\animations/condensation.html}.

%--------------------------------------------------------------------------------------------------------------------------------------------------------------------------------
\subsection{Effect of a magnetic field}
\label{sec:supp}

To illustrate the qualitative differences between a cloud's evolution with no ambient magnetic field, with the presence of a weak field, and a strong field, we show the distribution of cold gas within the domain at the end of the simulations at $t\approx 50$ Myr for simulations 2HN, 2HW, and 2HS in Figure~\ref{fig:rhocoldprojs}. In the non-magnetic simulation, the cloud expands symmetrically in the transverse directions, and a turbulent tail forms out of the  gas ablated from the cloud. This stream of gas is highly structured down to the smallest resolved scale, thus resulting in a large contact surface between the cloud gas and the ambient medium.

The structure of the cloud and its wake is remarkably different in the weak and strong magnetic field simulations, both compared to the pure HD case and to one another. In these cases, the wake is mostly confined within a thin plane along $y=0$. This tail is more extended in the other transverse (i.e. $z$) direction, a general effect seen in MHD cloud-wind simulations \citep[e.g.][]{gregori99,mccourt15,gronnow17}. Thus, the magnetic field breaks the symmetry of the initial conditions. In the weak field case, the wake is composed of four main filaments, rather than a single tail as in the strong field case (see the third panel from the top of Figure \ref{fig:rhocoldprojs}). In the strong field case, the cloud falls behind compared to the other cases (see the bottom two panels of Figure \ref{fig:rhocoldprojs}). As we show later, this is because of additional drag caused by the strong magnetic tension building up at the cloud's edge, which significantly slows down the cloud.

Figure~\ref{fig:bfieldslices} displays the magnetic field strength at the end of simulations 2HW and 2HS relative to the initial strength (i.e., the field amplification) along two mutually orthogonal slices across the domain. In the weak field case, the cloud `sweeps up' magnetic field, leading to the formation of a `draping layer' of significantly amplified magnetic field at the cloud's leading edge and around it. The magnetic pressure around the cloud is comparable to the ram pressure, indicating that field amplification is saturating \citep{jones96}. In the strong field case, the field does not fully drape around the cloud but rather bends due to the cloud moving at sub-Alfv\'{e}nic velocity, as discussed in Section \ref{sec:draping}. Field amplification is much less developed in the strong field case, the strongest field being mainly confined to the cloud's leading edge. In both cases there are also regions around the wake where the field is weaker than initially, although this is much more pronounced in the strong field case. However, the field is always amplified inside of the wake material itself and in its immediate vicinity.

%--------------------------------------------------------------------------------------------------------------------------------------------------------------------------------
\begin{figure}
\begin{center}
\includegraphics[height=0.9\textheight]{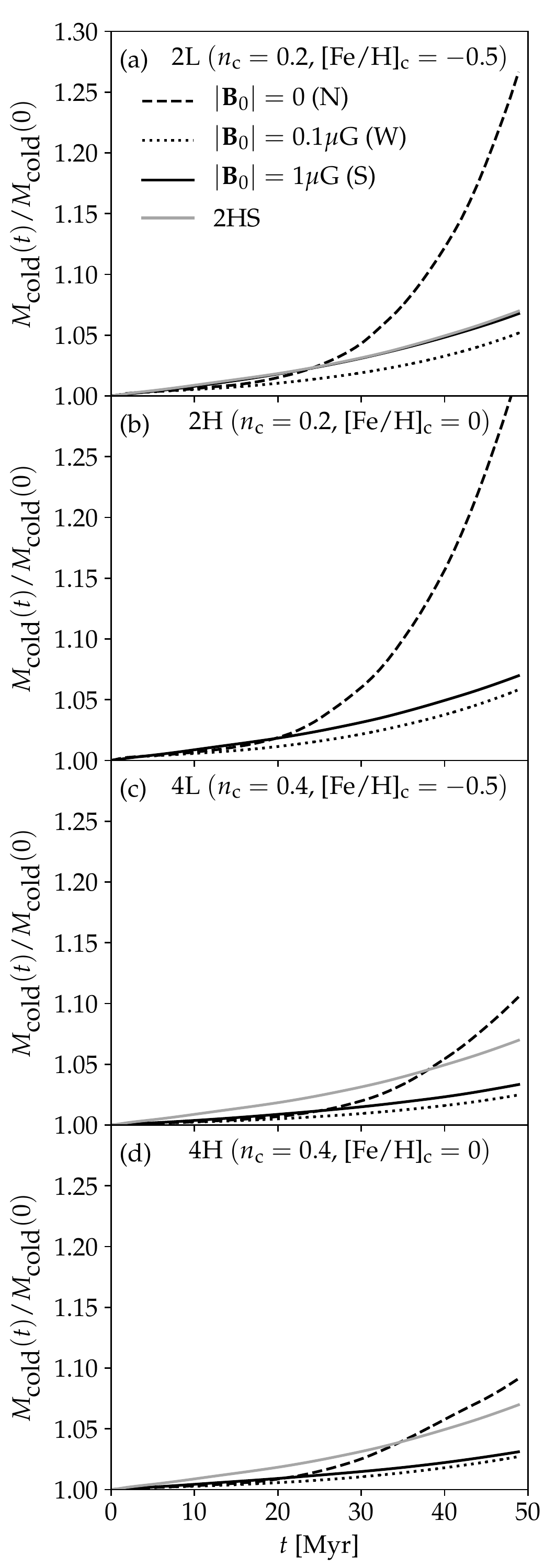}
\end{center}
\caption{Evolution of the gas condensation, quantified by the ratio of cold ($T < 5 \times 10^5$ K) gas mass relative to its initial value, in different simulations with an initial strong (solid), weak (dotted), and no (dashed) ambient magnetic field. Model 2HS from panel (b) is included in the other panels for comparison (gray).}
\label{fig:condensation}
\end{figure}
%--------------------------------------------------------------------------------------------------------------------------------------------------------------------------------

The evolution of the condensation along the wake of clouds with no ambient magnetic field, and with a weak and strong field, is shown in Figure~\ref{fig:condensation}. In either case, the amount of cold gas increases roughly linearly with time at $t \lesssim 20$ Myr. A substantial fraction of this cold gas mass is due to the cooling of cloud gas that initially has a temperature just above the threshold of $5 \times 10^5$ K. We know this because we have run a no-wind ($v_{\text{wind}}=0 \kms$) simulation where the condensation displays essentially the same behavior during this time frame (not shown). In other words, the condensation we see at $t \lesssim 20$ Myr is not caused by dynamical effects. In the pure HD case, the gas condensation rate increases substantially at later times, consistent with previous studies (\citetalias{marinacci10}; \cite{fraternali15}; \cite{armillotta16}). In striking contrast, {\em the gas condensation of cold gas along the wake of the cloud proceeds at a roughly linear rate throughout in the presence of an ambient magnetic field}. Importantly, this result is apparently not sensitive to the field's initial strength.

As a result, by $t \approx 50$ Myr, the overall mass of cold gas is significantly higher in the HD simulation compared to the simulations with an ambient magnetic field. Interestingly, the initially weaker ($\vert \mathbf{B}_0\vert=0.1 \mu$G) field suppresses the condensation slightly more than the initially stronger ($\vert \mathbf{B}_0\vert=1 \mu$G) field due to the draping of the weak field (see Section~\ref{sec:discussion}). Eventually, if the field strength is further decreased, the suppression of condensation starts to become less efficient and all results tend toward the hydrodynamical case as $\vert \mathbf{B}_0\vert \rightarrow 0 \mu$G, as expected. We have verified this by running a simulation with $\vert \mathbf{B}_0\vert=0.01 \mu$G (not shown), which we find to be almost indistinguishable from its $\vert \mathbf{B}_0\vert=0 \mu$G counterpart.

\begin{figure}
\begin{center}
\includegraphics[width=0.49\textwidth]{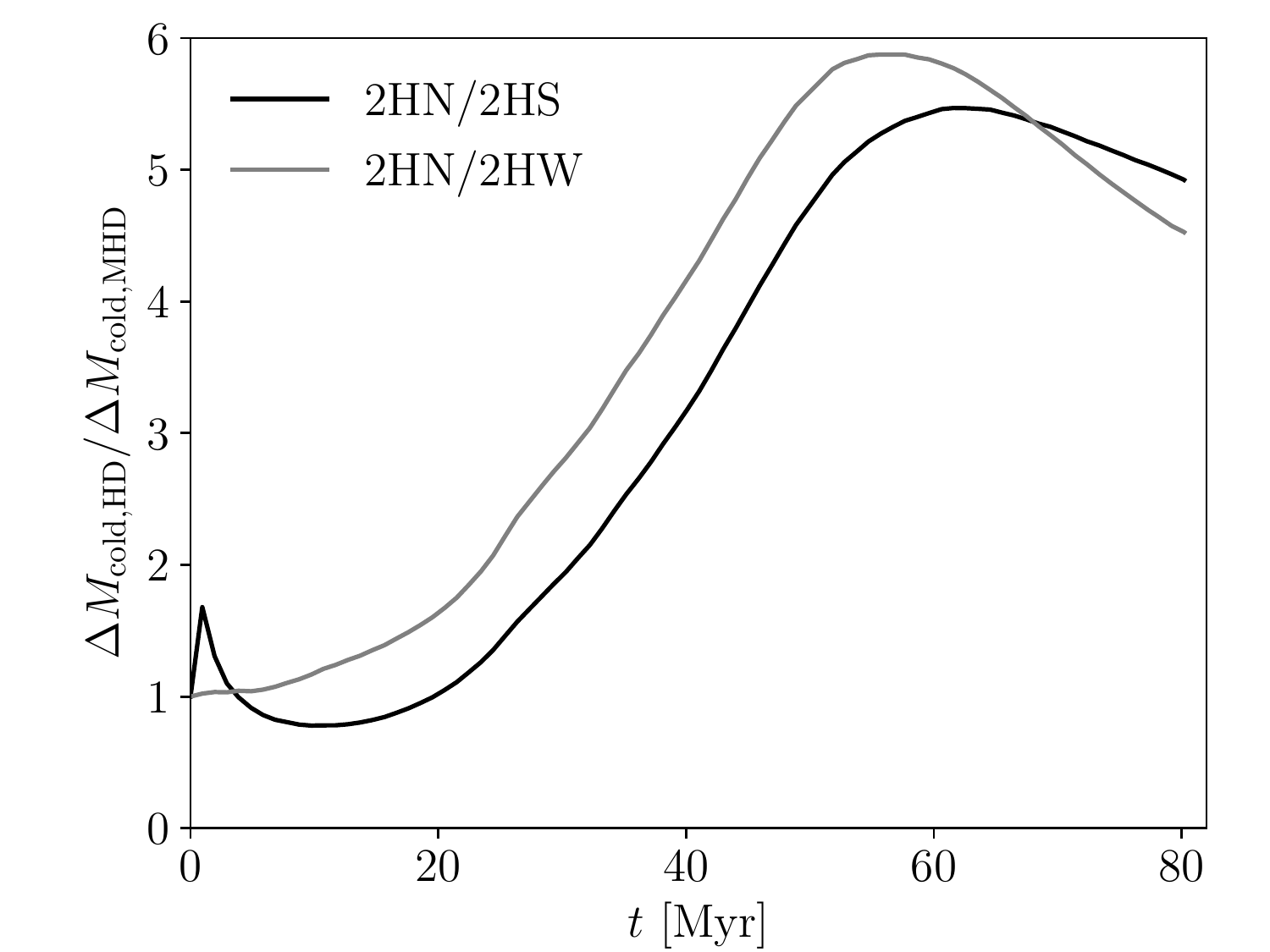}
\end{center}
\caption{Evolution of the ratio of the change in cold ($T < 5 \times 10^5$ K) gas mass since $t=0$ Myr between the non-magnetic simulation 2HN and the two equivalent magnetic simulations 2HW (gray) and 2HS (black). This is essentially the factor that the condensation is being suppressed with by the magnetic field. Note that the evolution is shown until $t\approx 80$ Myr.}
\label{fig:hdmhdratio}
\end{figure}

The typical travel time of Galactic fountain clouds is about 80 Myr \citep{marasco13}. However, we generally run our simulations only to $t\approx 50$ Myr to keep computational costs feasible. To probe the later evolution, we run a small set of simulations (2HN, 2HW, and 2HS) to $t\approx 80$ Myr. For $t \gtrsim 55$ Myr, the growth of cold gas mass slows down in simulation 2HN, and the trend becomes approximately linear, albeit with quite a steep slope. At $t \approx 80$ Myr, the amount of cold gas has almost doubled since $t=0$ Myr. We compare this to the full evolution of cold gas mass in the corresponding MHD simulations in Figure \ref{fig:hdmhdratio}. Rather than repeating panel (b) in Figure \ref{fig:condensation} with the later evolution included, we show the comparison in a complementary way by plotting the ratio of the change in cold gas mass in simulation 2HN to the two MHD simulations. This essentially is the factor by which the condensation is being suppressed with by the magnetic field. As can be seen, the magnitude of this suppression factor keeps growing until $t\approx$ 55(60) Myr, reaching a maximum of about 5.5(6) for the strong(weak) field simulation. After this, it starts to fall off but is still above 4 by $t\approx 80$ Myr. The decrease in suppression is caused by the amount of condensation growing linearly at late times in the non-magnetic case, but slightly superlinearly in the two magnetic cases. The amount of cold gas starts to grow faster in the strong field case (2HS), overtaking the weak field case (2HW) at $t \approx 70$ Myr. In summary, although there is a change in the cold gas mass trends after 50 Myr, in this case the suppression remains significant. The first 50 Myr of the evolution of the cold gas mass suppression follows similar trends for the other simulations, but with different normalizations. In all these cases, the suppression factor is still growing at 50 Myr, with values ranging from about 2.5 for simulation 4HS to about 5 for simulation 2LW.

\begin{figure}
\begin{center}
\includegraphics[width=0.49\textwidth]{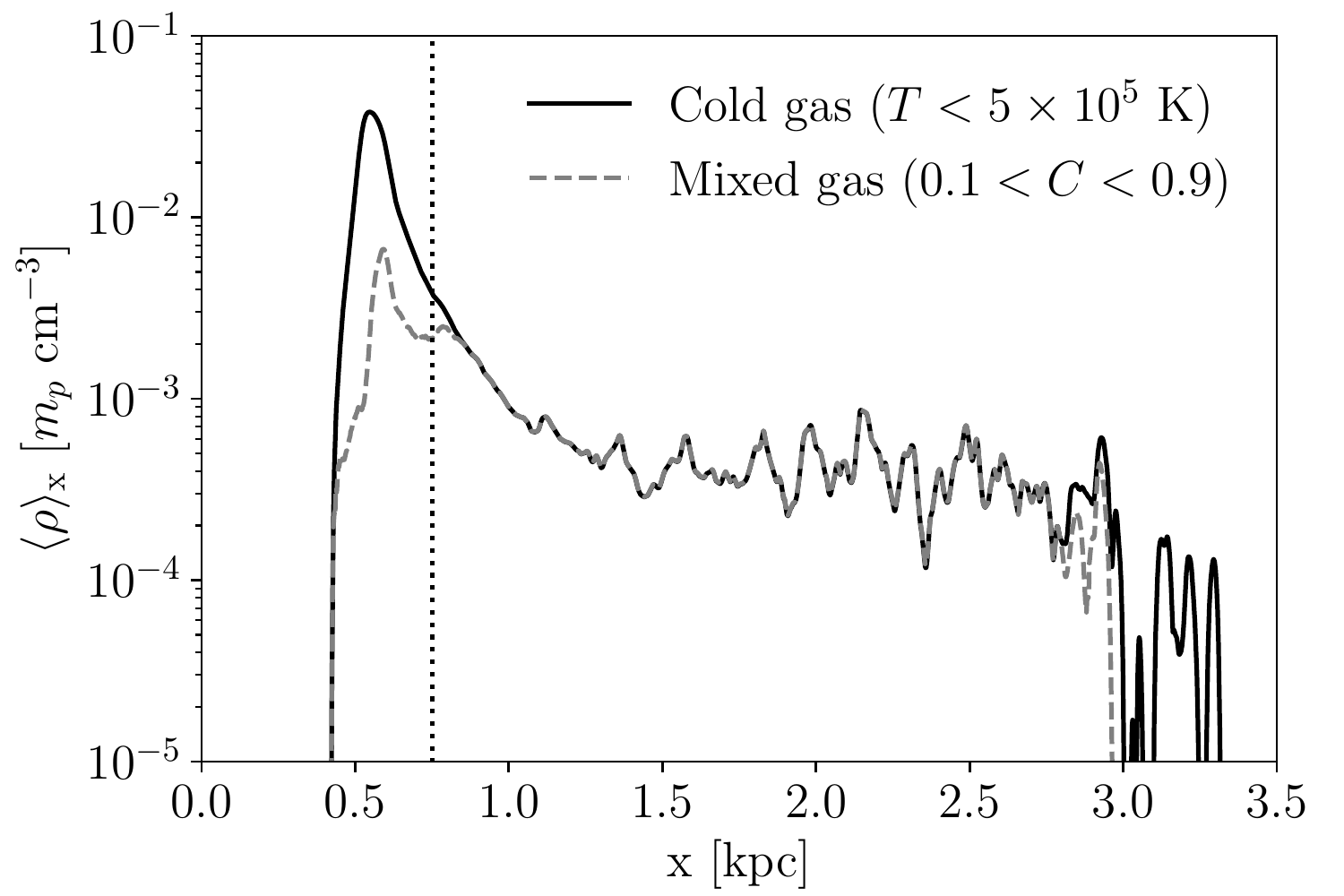}
\end{center}
\caption{Mean density of $y \in [-0.4,0.4], z \in [-0.4,0.4]$ slices along $x$ of cold gas (solid black) and mixed gas (dashed gray) in simulation 2HS at $t\approx 50$ Myr. The dotted vertical line marks approximately the transition from cloud to wake, defined as the point where the width of the cloud stops shrinking (see Figure \ref{fig:rhocoldprojs}).}
\label{fig:coldmixgas}
\end{figure}

We find that the main process driving gas condensation along a cloud's wake is the mixing between gas ablated from the cloud and halo gas. We demonstrate this in Figure~\ref{fig:coldmixgas} which shows the mean density of $y \in [-0.4,0.4], z \in [-0.4,0.4]$ slices along $x$ of cold gas and mixed gas at the end of simulation 2HS. We quantify the degree of mixing with aid of the color tracer $C$; cells with $0.1 < C < 0.9$ are considered to be composed of a mixture of cloud and halo gas. Apparently, the mean density of mixed gas closely follows the density of cold gas throughout the wake. The large difference to the left of the dotted line is the remains of the cloud core which has not mixed. The results are essentially the same for other simulations (not shown).

%--------------------------------------------------------------------------------------------------------------------------------------------------------------------------------
%--------------------------------------------------------------------------------------------------------------------------------------------------------------------------------

In summary, we find that {\em the presence of a magnetic field significantly reduces the amount of gas condensation along the wake of clouds moving through the halo}. We believe that the reason behind this is that magnetic fields enhance the stability of a cloud against ablation and eventual break-up due to hydrodynamical instabilities \citep[mainly KH; e.g.][]{banda-barragan16}, thus effectively reducing the contact surface between halo and cloud material (at the limiting resolution), and consequently the amount of gas mixing between the two phases (see Section \ref{sec:instabilities}). It is worth emphasizing that condensation still occurs in all cases, as can be seen by the mass of cold gas always increasing monotonically in Figure \ref{fig:condensation}. However, the reduction in condensed gas mass caused by the magnetic field improves agreement with observational constraints on Galactic fountain accretion rates (see Section \ref{sec:comparisonobs}).

%--------------------------------------------------------------------------------------------------------------------------------------------------------------------------------
%--------------------------------------------------------------------------------------------------------------------------------------------------------------------------------

\subsubsection{Effect of cloud density}
The effect of varying the initial cloud density on the amount of condensation can be seen by comparing panels (a) and (c), and panels (b) and (d), displayed in Figure~\ref{fig:condensation}.
Qualitatively, there is little difference between the condensation fraction of diffuse and dense clouds. In absolute terms, the amount of condensation is generally lower in the case of dense clouds. In the pure hydrodynamic case, this can easily be understood as a result of the dependence of the KH instability timescale on the (average) cloud density when $\rho_c \gg \rho_h$: $t_{\text{KH}} \propto \sqrt{\rho_c}$ (see eq.~\ref{eq:khtimescale}). In fact, if we express the evolution of the condensation of denser clouds in terms of a rescaled time parameter given by $t' = \left( t_{\text{KH},n=0.4}/t_{\text{KH},n=0.2} \right)~t=\sqrt{2}t$, then most of the difference between dense and diffuse clouds goes away in the HD and weak field cases. In the presence of a strong magnetic field, some, but far from all, of the difference disappears in terms of this rescaled time parameter. It is interesting to note that even the \emph{absolute} (rather than relative to the initial cold gas mass) amount of condensation (i.e., $M_{\text{cold}}(t)-M_{\text{cold}}(0)$), is slightly lower for the high density clouds as well. Thus, the (perhaps na\"\i{}ve) expectation that an initially higher amount of cold gas available should lead to more mixing and condensation turns out to be incorrect, because, in fact, denser (and more massive) clouds are more stable compared to more diffuse clouds as a result of the higher cloud : halo density contrast.

In hindsight, the role of the initial cloud density can be easily understood because it enters the problem essentially as a simple scaling parameter, at least in the case of weak or vanishing magnetic fields.

\subsubsection{Effect of metallicity}
\label{sec:metal}

It is well-known that the cooling efficiency of gas increases with its metallicity. Even at gas metallicities of order [Fe/H] $\sim -1$, typical
of HVCs and the MS \citep{fox14}, the metal-line cooling is at least as important as recombination emission from ionized hydrogen
\citep{bland-hawthorn13}. Nevertheless, the net effect of the metallicity of the cloud or the ambient medium on the overall condensation of gas is not obvious beforehand. Increased cooling along the cloud's wake should obviously increase the amount of cold gas. At the same time, increased cooling around the cloud will generally suppress cloud ablation \citep{cooper09}, thus making less cloud material available to mix with the halo gas in the first place.

We assess the overall effect of metallicity on gas condensation by comparing runs that are identical except for the cloud's initial metallicity. Consider, for example, the gas condensation along the wake of clouds shown in panels (a) and (b) in Figure~\ref{fig:condensation}. These clouds have the same initial density but initial gas metallicities that differ by a factor of about three. The same is true for the runs corresponding to panels (c) and (d) in the same figure, only that the cloud's initial density is a factor of two higher compared to the clouds corresponding to panels (a) and (b). In the absence of an ambient magnetic field, a higher initial metallicity results in slightly more condensation for low-density clouds, relative to the initial cold gas mass. However, the opposite is true for the high density clouds. In this case, the ablated gas is dense enough to cool efficiently, even at lower metallicity. Therefore, the increased cooling efficiency that results from a higher metallicity does not lead to additional significant condensation in the wake. Rather, the stabilizing effect of the additional cooling on the cloud apparently becomes dominant, thus resulting in less mixing overall. In the presence of an initially weak magnetic field, clouds with higher metallicity experience slightly more condensation along their wakes, regardless of their density. But the effect is much smaller compared to the pure HD case. Interestingly, in the presence of a strong magnetic field, the overall amount of condensation is insensitive to the cloud's metallicity. This suggests that when there is a magnetic field, the mixed gas is already cooling faster than new gas is being stripped, so that further reducing the cooling timescale by increasing metallicity has no significant effect.

%--------------------------------------------------------------------------------------------------------------------------------------------------------------------------------
\begin{figure}[h]
\begin{center}
\includegraphics[width=0.49\textwidth]{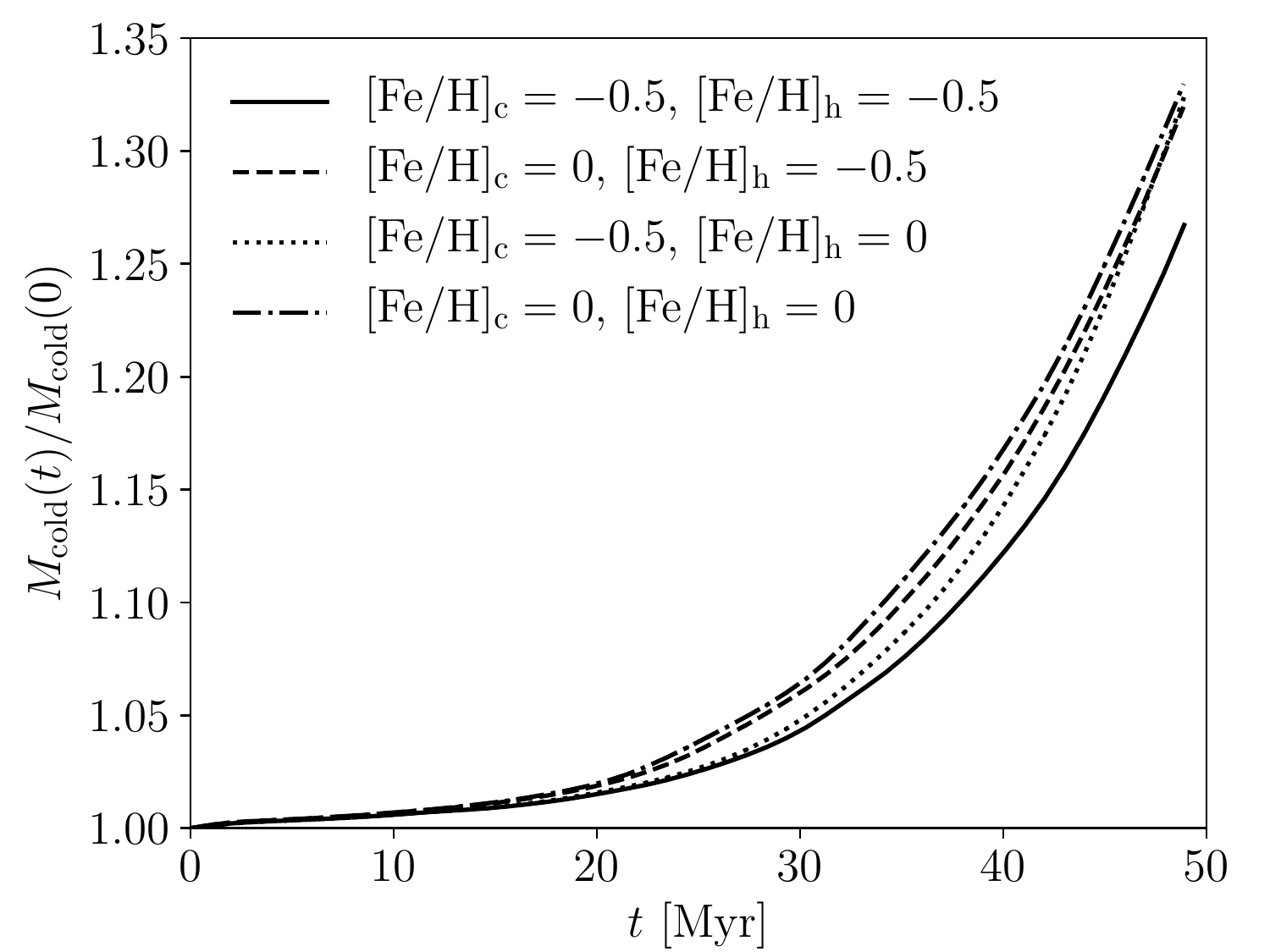}
\end{center}
\caption{Gas condensation for low-density clouds in pure HD simulations, for different initial cloud and halo metallicities. }
\label{fig:condensation_metallicity}
\end{figure}
%--------------------------------------------------------------------------------------------------------------------------------------------------------------------------------

We assess the effect of metallicity on gas condensation in more detail by running two simulations similar to 2LN and 2HN, except that the corona has an initial solar metallicity (rather than [Fe/H]$=-0.5$). In one case (2LHN), the halo has higher metallicity than the cloud; in the other (2HHN), both halo and cloud have the same (solar) metallicity. The result of this exercise, along with the results corresponding to 2LN and 2HN, are displayed in Figure~\ref{fig:condensation_metallicity}. Perhaps as expected, we find that the higher the metallicity of either the cloud or the halo, the more gas condensation, although the effect is relatively modest in all cases. The highest amount of condensation is obtained in the case where both halo and cloud have the same high metallicity, in agreement with \citetalias{marinacci10}.

\subsection{Effect of cloud velocity}
\label{sect:highvel}

%--------------------------------------------------------------------------------------------------------------------------------------------------------------------------------
\begin{figure}[h]
\begin{center}
\includegraphics[width=0.49\textwidth]{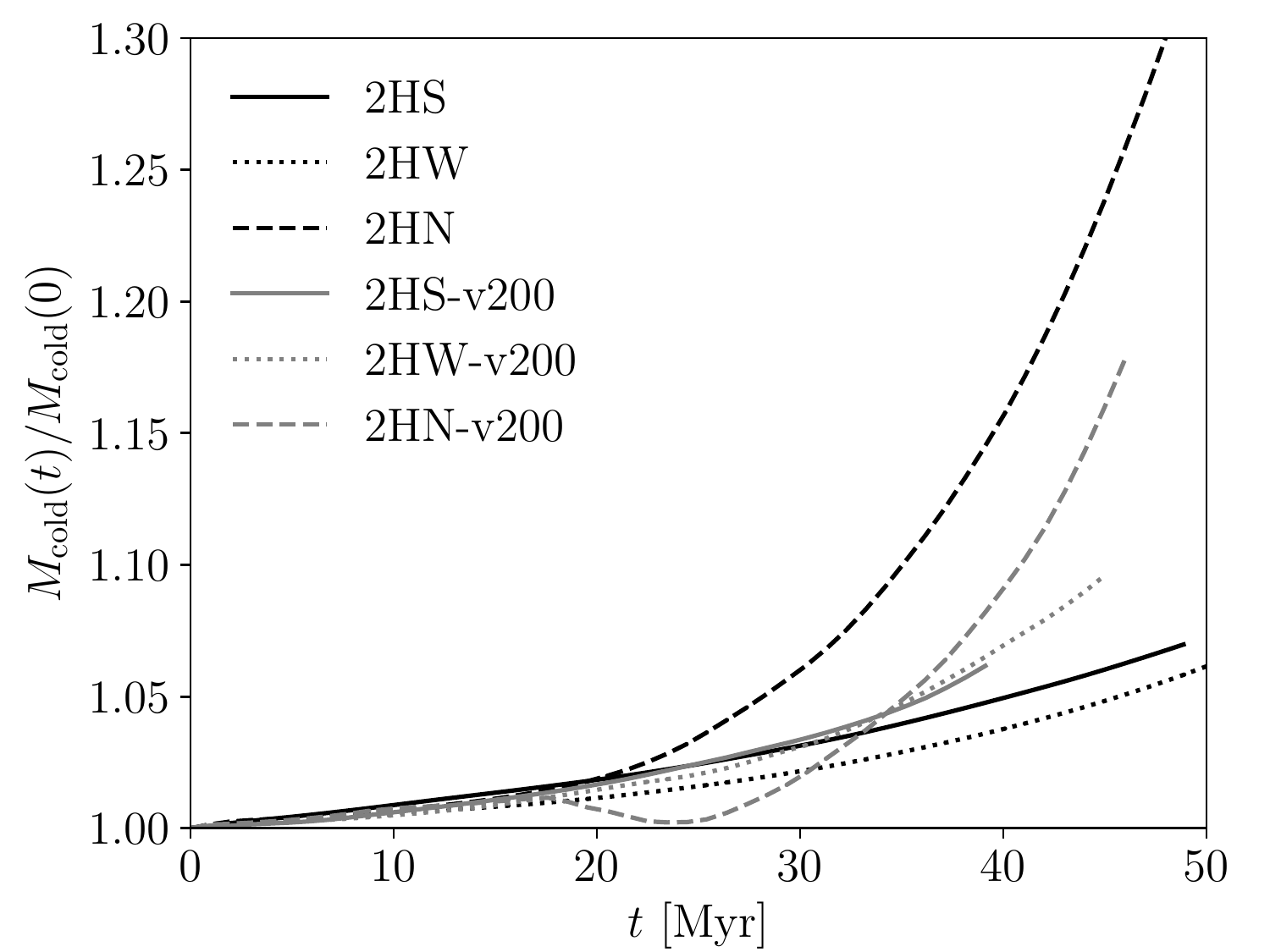}
\end{center}
\caption{Gas condensation for simulations 2HN, 2HW, and 2HS in the subsonic regime ($v_{\text{wind}}=75 \kms$, $\mathcal{M}=0.45$) and their transonic counterparts 2HN-v200, 2HW-v200, and 2HS-v200 ($v_{\text{wind}}=200 \kms$, $\mathcal{M}=1.2$). Note that the high-velocity magnetic simulations are barely distinguishable from their lower velocity counterparts at early times.
}
\label{fig:velocity}
\end{figure}
%--------------------------------------------------------------------------------------------------------------------------------------------------------------------------------

All the simulations described so far have been in the subsonic regime with $\mathcal{M}=v_{\text{wind}}/v_c=0.45$. We might expect that increasing the velocity will simply increase the amount of condensation based on the KH timescale given in Section \ref{sec:instabilities}. However, as we go from the subsonic to supersonic regimes, the assumption of incompressibility used to derive this timescale becomes increasingly inapplicable. Also, as we describe later, the behavior is complicated by the additional heating and the increased importance of the Rayleigh-Taylor (RT) instability at higher velocities. To examine how the cloud velocity affects the condensation, we run three simulations corresponding to 2HN, 2HW, and 2HS with a higher velocity of $v_{\text{wind}}=200 \kms$ representative of an HVC. These are in the transonic regime ($\mathcal{M}=1.2$). They are not run to $t=50$ Myr, due to being numerically expensive. We compare the cold gas mass evolution for these simulations with their subsonic counterparts in Figure~\ref{fig:velocity}.

The non-magnetic, transonic simulation (2HN-v200), in stark contrast to all the other runs, shows significant deviation from a monotonic trend in the build-up of cold gas mass. While it follows the trend of simulation 2HN at early times, it shows a {\em decrease} in the amount cold gas between $t=15$ Myr and $t=24$ Myr. However, the cold gas mass fraction subsequently increases again, beyond the condensation level seen in the magnetic simulations. There are a number of reasons behind this perhaps unexpected behavior. First, the higher cloud speed leads to additional adiabatic heating from stronger compression, counteracting the cooling of the mixed gas. Second, at this higher speed, the RT instability also becomes significant within the timescale of the simulation. While the increased velocity leads to additional stripping caused by the KH, and later on RT, instabilities this does not lead to a corresponding increase in the overall cooling efficiency. This is due to the stripped gas being dispersed over a larger region and so becoming diluted compared to lower velocity simulations. Because of the increased stripping and stronger shock, the cloud is mostly destroyed by the end of the simulation, having dispersed into small filaments and cloudlets (at the limiting resolution). We note that the high Mach number makes the simulation more numerically challenging than our other subsonic simulations. In particular, at later times the cloud is affected by numerical grid alignment artifacts, which cause its shape to become unnaturally boxy with a pinlike protrusion at its leading edge.

Transonic clouds moving through a highly magnetized medium (2HS-v200) experience a significant field amplification around them, which leads to slightly less condensation than at low velocity as in the weak field case (see Section~\ref{sec:draping}) at early times. This is a consequence of the cloud being in the super-Alfv\'{e}nic regime, which also causes the wake to have a twin tail morphology resembling the weak field case. In contrast, at $t \gtrsim 30$ Myr the cold gas mass overtakes that of the lower velocity counterpart, possibly due to an increased amplitude of the RT instability in the $z$ direction. However, these differences are generally not significant, and the mass of cold gas mostly follows the subsonic case too closely to be discernible in the figure.

In the presence of a weak ambient field, the condensation along the wake of a transonic cloud is higher at all times compared to the subsonic case. The magnetic field around a transonic cloud is not amplified significantly more than in the case of a subsonic cloud. Overall, the higher speed leads to a stronger ablation (through the KH instability) and thus to an increased mixing.

We conclude that \emph{the amount of condensation, while still positive, is systematically lower in the presence of an ambient magnetic field, regardless of the cloud's speed.}

\subsection{Effect of magnetic field orientation}
\label{sec:oblique}

Our choice of an ambient magnetic field that is initially perfectly aligned with the plane transverse to the wind's direction is admittedly arbitrary. It is therefore important to assess the robustness of our results with respect to the assumed field orientation.

%--------------------------------------------------------------------------------------------------------------------------------------------------------------------------------
\begin{figure}[h]
\begin{center}
\includegraphics[width=0.49\textwidth]{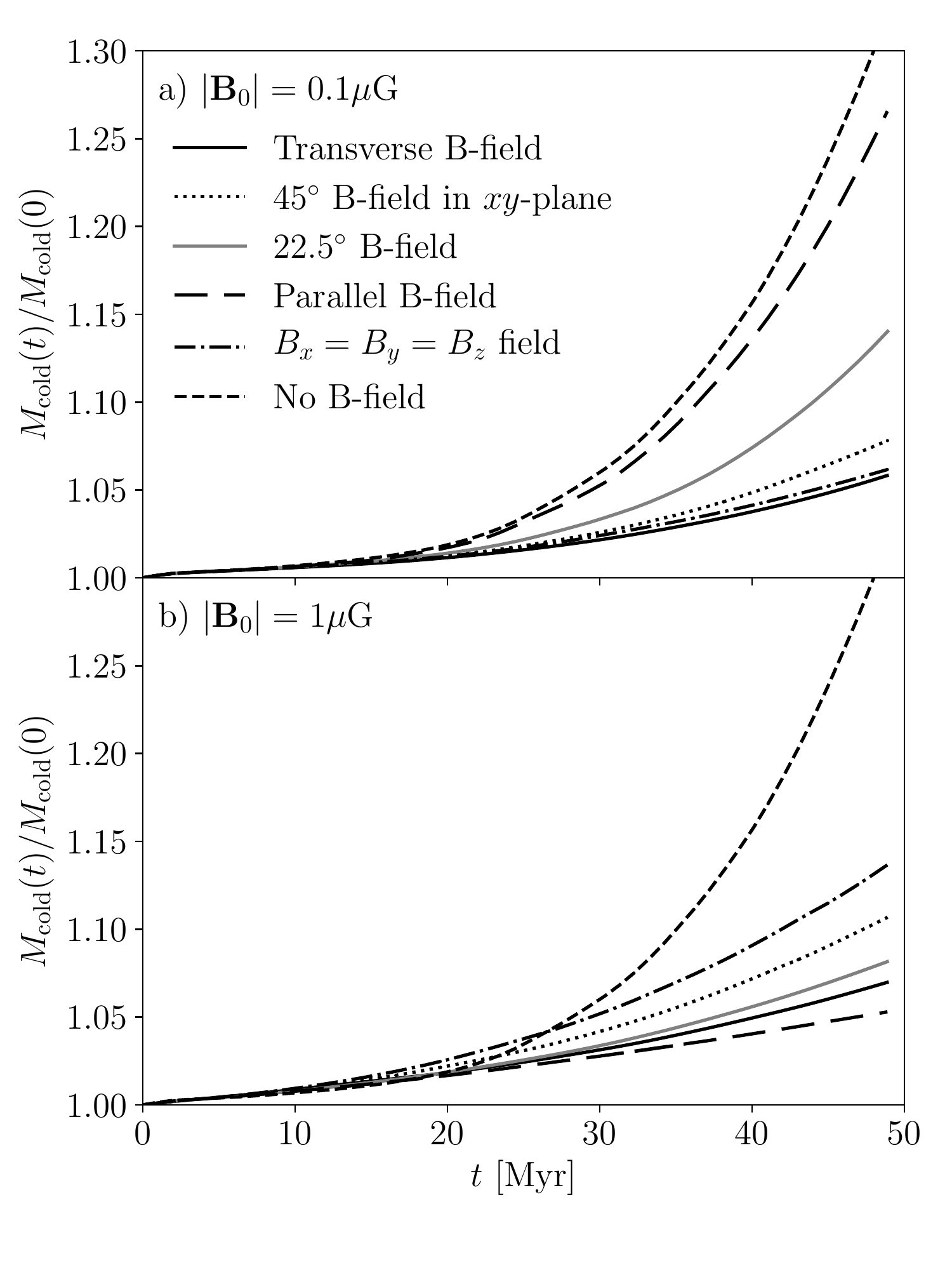}
\end{center}
\caption{Panel (a) shows gas condensation for simulation 2HW (solid black), which has a transverse magnetic field; 2HW-45 (dotted) which has a field at 45$^\circ$ in the $xy$-plane; 2HW-225 (solid gray), which has a field at 22$\stackrel{\hbox{$\scriptstyle{\circ}$}}{\hbox{.}}$\hspace{-0.6pt}5 in the $xy$-plane; 2HW-par (long dashed), which has a magnetic field parallel to the direction of motion; 2HW-eq (dash dotted), which has an oblique field with equal $x$, $y$, and $z$ components; and 2HN (short dashed), which has no magnetic field. Panel (b) shows the same field orientations but for the strong field simulations.}
\label{fig:fieldorientation}
\end{figure}
%--------------------------------------------------------------------------------------------------------------------------------------------------------------------------------

To this end, we run simulations with initial conditions identical to 2HW and 2HS, except that for these the magnetic field is (i) parallel to the velocity ($\mathbf{B}=\left(B_0,0,0\right)$); (ii) at a 45$^\circ$ angle in the $xy$-plane ($\mathbf{B}=\frac{B_0}{\sqrt{2}} \left(1,1,0\right)$); (iii) at a 22$\stackrel{\hbox{$\scriptstyle{\circ}$}}{\hbox{.}}$\hspace{-0.1pt}5 angle in the $xy$-plane ($\mathbf{B}=B_0\left(0.92,0.38,0\right)$); and (iv) with all components being equal ($\mathbf{B}= \frac{B_0}{\sqrt{3}} \left(1,1,1\right)$). The amount of condensation in these simulations is compared to that of the usual transverse field and the non-magnetic case in Figure~\ref{fig:fieldorientation}.

Clearly, the field orientation has a strong impact on the amount of condensation, depending on the field strength. Consider, for example, the case of clouds in a weak magnetized medium. When the field is parallel to the flow, it is barely amplified and thus has a negligible effect. The 45$^\circ$ field and the field with equal components, however, are effectively draped around the cloud (see Section \ref{sec:draping}). The case where the field initially has equal components is nearly indistinguishable from the transverse field case, in agreement with \cite{banda-barragan16}. Due to the symmetry in the initial conditions, the angle in the transverse (i.e. $yz$) plane is irrelevant. The important quantity is the magnitude of $\mathbf{B}$ projected onto this plane (i.e., $\vert\mathbf{B}\vert_{yz}^2=B_y^2+B_z^2$). The difference between the case of an initial 45$^\circ$ field and a field with all components equal to one another can thus be explained by the former having a transverse length of $\vert \mathbf{B}\vert_{yz}=\sqrt{1/2}$ while the latter has $\vert \mathbf{B}\vert_{yz}=\sqrt{2/3}$. The 22$\stackrel{\hbox{$\scriptstyle{\circ}$}}{\hbox{.}}$\hspace{-0.1pt}5 case is intermediate between the 45$^\circ$ and parallel cases, and so is the amount of condensation seen in this run, as expected.

In contrast, when a cloud is moving through a strongly magnetized medium, the magnetic field does not fully drape around the cloud but rather bends akin to a shock bow. The initially strong field is essentially only amplified at the leading edge of the cloud and only by a small factor of about 2. Thus, the amplification is generally not driving the dependence on field orientation for the suppression of condensation in this case. In contrast to the weak field case, the parallel field is actually slightly more effective at suppressing condensation than the transverse field. At an interface, only the component of the field that points along the interface has an effect on the KH instability there. Thus, one possible explanation is that the transverse field, only bending rather than draping, is generally less aligned with the cloud-wind interface and, especially, the tail-wind interface compared to the parallel field. Another possibility is that the RT instability becomes important in the strong field case, and the amplification of the field limited to the front of the cloud enhances this instability (see Section \ref{sec:instabilities}). For the parallel field, the cloud has two thin tails, rather than one, at the top and bottom of the cloud in the $y$ direction. The 45$^\circ$ and equal components fields show significantly more condensation. The strong field pulls the cloud and bends the tail such that the tail becomes fully exposed to the wind, unlike all the other simulations where the tail is partially shielded by the cloud being directly upstream. This causes the tail to spread out much more than in the other cases. The increase in surface area increases the mixing and in turn the amount of condensation. In the 22$\stackrel{\hbox{$\scriptstyle{\circ}$}}{\hbox{.}}$\hspace{-0.1pt}5 case the same effect is present but to a lesser degree, leading it to lie between the 45$^\circ$ and parallel cases, as would be expected.

In short, clouds moving in a weak field are in the highly super-Alfv\'{e}nic regime, and thus experience magnetic field draping (see Section~\ref{sec:draping}). The key point is that field amplification requires draping, which depends on the field orientation. As we have discussed, the higher the field amplification, the stronger hydrodynamical instabilities are suppressed, which leads to less ablation, less mixing, and thus to less condensation. Clouds in the strong field are in the sub-Alfv\'{e}nic regime, and the field draping is effectively absent. However, amplification is not necessary for the field to be dynamically important in this case.

In summary, while the precise amount of condensation apparently depends on the structure and strength of the ambient magnetic field, {\em gas condensation is generally significantly suppressed whenever there is an ambient magnetic field present}, compared to identical clouds moving through an non-magnetized medium. Similarly to the transverse magnetic field, however, condensation does still occur in all cases, with the mass of cold gas monotonically increasing.

\subsection{Effect of resolution}
\label{sec:resolution}

We denote spatial resolution as $\mathcal{R}_{N}$, where $N=r_c/\Delta x$ (i.e., the number of cells per cloud radius). Our standard simulations have six levels of refinement corresponding to a maximum resolution of $\mathcal{R}_{64}$. Such a limiting resolution has been found to be sufficient, in the sense that results in {\em adiabatic} MHD cloud-wind simulations adopting this resolution appear to be converged \citep{banda-barragan18}. However, the minimum spatial scale to properly account for the effect of radiative cooling is generally smaller than the one required to resolve magnetic fields (see Section~\ref{sec:comparison}).

\begin{figure}[h]
\begin{center}
\includegraphics[width=0.49\textwidth]{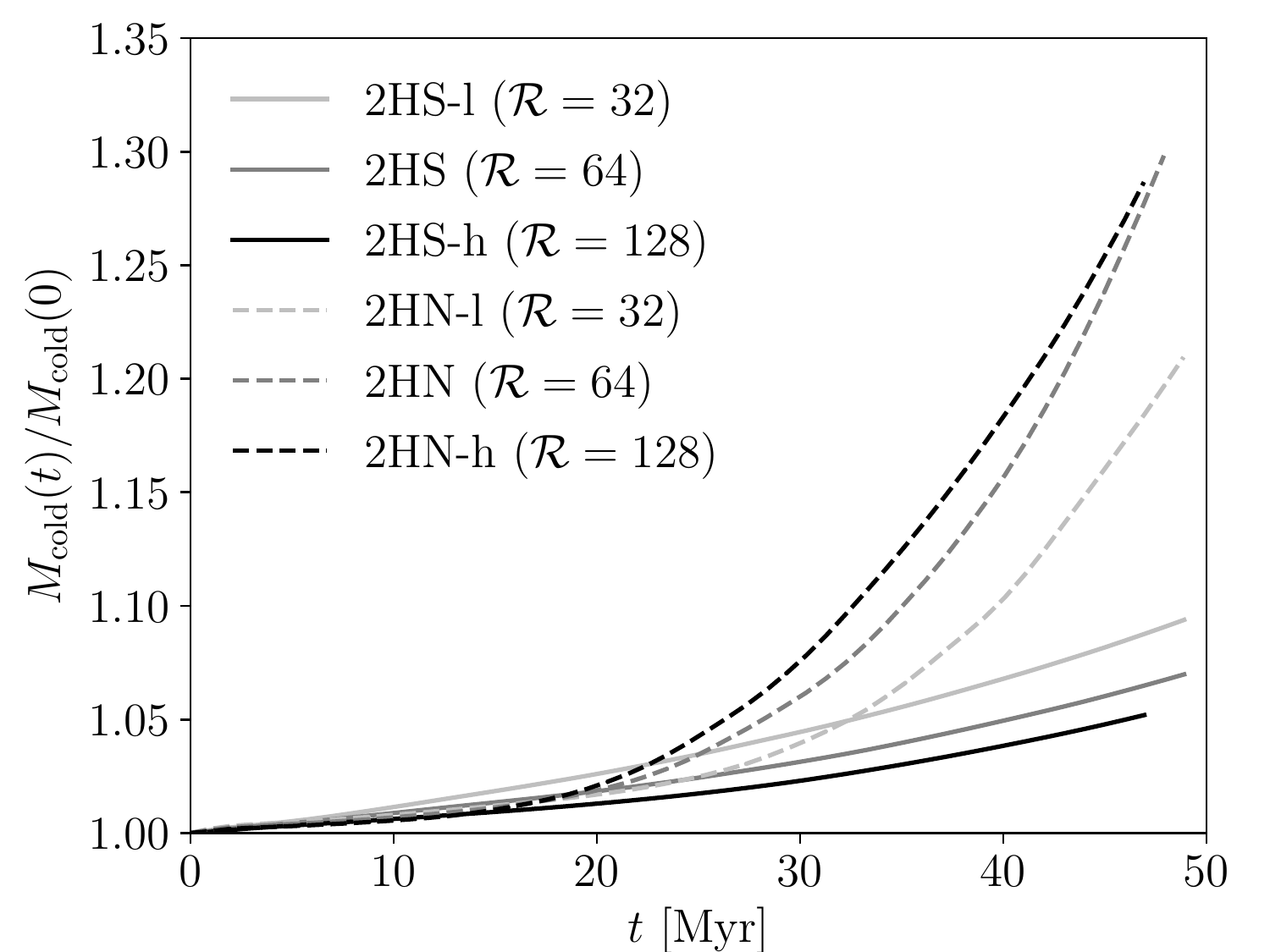}
\end{center}
\caption{Gas condensation for simulation 2HS (solid) and simulation 2HN (dashed) at low ($\mathcal{R}_4=32$ cells/$r_c$, light gray), medium (standard; $\mathcal{R}_5=64$ cells/$r_c$, dark gray), and high ($\mathcal{R}_6=128$ cells/$r_c$, black) spatial resolution.}
\label{fig:mhdvshd_res}
\end{figure}

We compare the amount of gas condensation in simulations 2HS and 2HN at our standard resolution with simulations that have half and twice the resolution (i.e. $\mathcal{R}_{32}$ and $\mathcal{R}_{128}$, respectively) but are otherwise identical to their counterpart. The results are displayed in Figure~\ref{fig:mhdvshd_res}. Note that the low (high) resolution runs contain one less (one additional) refinement level. For the high-resolution runs, we also adopt a slightly lower threshold value for the second derivative error norm of the density, which is used as a criterion for refinement (see Section~\ref{sect:ICs}).

As can be seen, the amount of gas condensation varies with resolution, and depending on whether an ambient magnetic field is present. In its absence, the amount of condensation generally {\em increases} with resolution. The reason for this is that increasing the limiting spatial resolution allows us to resolve KH perturbation at progressively smaller scales, which leads to more stripping and mixing \citep[e.g.][]{scannapieco15}.

The cold gas mass in the high-resolution simulation starts to grow less steeply at late times compared to at standard resolution. This results in the cold gas masses being about equal by the end of the simulations. However, this relative decrease in the slope of the cold gas mass at $t \gtrsim 35$ Myr coincides with the cloud in the high-resolution run becoming noticeably affected by a numerical grid alignment artifact associated with the mesh refinement. The cloud appears to `snap to' the underlying base level grid acquiring a box-like overall shape as a result.

When a magnetic field is present, the amount of condensation {\em decreases} with resolution substantially, by about 40\% by $t=30$ Myr. This is because the tail, which contains most of the condensed gas, is not resolved at any of our resolutions. This can be inferred from the fact that the tail is about 5 cells wide in the $y$ direction at all resolutions. The physical width of the tail thus decreases with resolution because it is proportional to the cell width. Its narrowness appears to be limited by the 3-point stencil width of the interpolation scheme used in the simulations. Thus, the volume of the tail is overestimated, leading to too much mixing. To summarize, the cold gas mass decreases with resolution for the magnetic case but increases with resolution in the non-magnetic case. Therefore the relative difference between the two cases (i.e., the suppression of condensation by the magnetic field) increases with resolution.

The quantitative differences in the amount of condensation that results in simulations with different (increasing) spatial resolution indicates that our results are not converged. However, it is important to note that there is no qualitative difference in the resulting amount of condensed gas across runs with varying resolution. Indeed, the trends in the evolution of the gas condensation are overall comparable. Most crucially, the effect of the magnetic field in suppressing the condensation of gas increases with resolution. We interpret this as an indication that this effect is not a numerical artifact caused by poor spatial resolution, and is thus significant.

%--------------------------------------------------------------------------------------------------------------------------------------------------------------------------------
%--------------------------------------------------------------------------------------------------------------------------------------------------------------------------------

\subsection{Effect of AMR and hyperbolic divergence cleaning}
\label{sec:amr}
Running our wide array of high-resolution simulations is only computationally feasible through the use of AMR. But care has to be taken when choosing the refinement criteria, because there is no universal strategy and it rather has to be adopted according to the problem at hand. Our choice of refining the grid depending on the density gradient, as well as the adopted threshold value, are somewhat arbitrary. It is therefore important to assess whether our results may be affected by these choices.

Using an adaptive grid also forces us to use a divergence cleaning scheme to minimize divergence in the magnetic field, which is supposed to be solenoidal. When a static grid is used, {\sc Pluto} provides an alternative method known as constrained transport \citep[CT;][]{evans88}. The CT procedure ensures that $\nabla \cdot \mathbf{B}$ is conserved to machine precision and thereby that $\nabla \cdot \mathbf{B}=0$ to similar precision because our initial magnetic field is divergence free, by construction. The downside is that the traditional implementation of CT, as used in {\sc Pluto}, is not feasible for adaptive meshes, as it relies on staggered grids. This is why we use a divergence cleaning scheme instead in all of our AMR simulations. Thus, by comparing static grid simulations using divergence cleaning to ones using CT, known to have no divergence, we can check if the divergence cleaning scheme is sufficient for our purposes.

Thus, we compare simulations 2HS and 2HN to runs with initial conditions, which are identical in every aspect --- labeled 2HS-s and 2HN-s, respectively --- except that these are run using a static (rather than an adaptive, dynamic) grid. We follow the same approach and compare the low resolution ($\mathcal{R}_{32}$) simulations 2HS-l and 2HN-l to static grid versions (labeled 2HN-ls and 2HS-ls, respectively). We also run two simulations at low and standard resolution, 2HS-ls-ct and 2HS-s-ct, which in addition to having a static grid use CT instead of divergence cleaning. Because the static grid runs are very computationally expensive, we have to use a significantly smaller simulation domain for these to be feasible. For the non-magnetic simulations, this is not an issue, as the wake does not extend beyond $x\approx 2.5$ kpc even at the end of the simulation. However, the other simulations (2HS-ls, 2HS-ls-ct, 2HS-s, and 2HS-s-ct) are limited by the smaller volume as the cloud moves farther along the $+x$ direction and its tail extends significantly further in the strong field case (see Figure~\ref{fig:rhocoldprojs}). As a consequence, we cannot run them beyond about 25 Myr, at which point some fraction of the cold gas starts to flow out through the $+x$ boundary. In addition, we were only able to run simulation 2HS-ls to around $t=18$ Myr, after which the simulation becomes unstable. Thus, we can only compare the first half of the cold gas mass evolution between static and adaptive grids in the magnetic cases.

\begin{figure}[h]
\begin{center}
\includegraphics[width=0.49\textwidth]{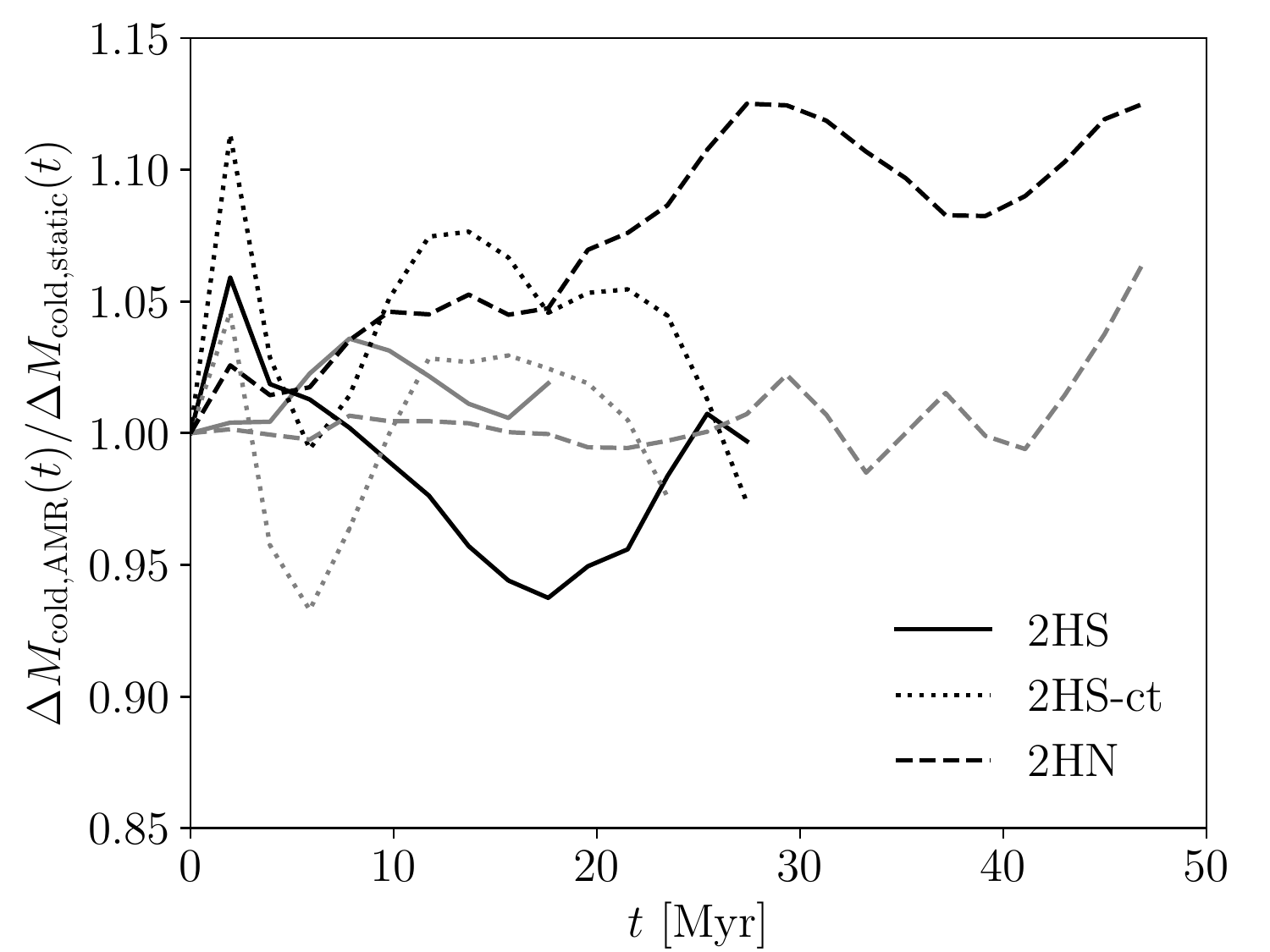}
\end{center}
\caption{Evolution of the gas condensation in AMR simulations relative to their corresponding static grid counterparts: 2HS compared to a static grid simulation using the same divergence cleaning scheme (solid), 2HS compared to a static grid simulation using constrained transport (dotted), and 2HN (dashed), at low (gray) and our standard (black) resolution. Note that $\Delta M_{\text{cold}}$ is used rather than $M_{\text{cold}}$ (i.e., the initial mass of the cloud is subtracted, to emphasize the differences).}
\label{fig:staticgrid}
\end{figure}

Figure~\ref{fig:staticgrid} shows the difference in the cold gas mass since $t=0$, $\Delta M_{\text{cold}}(t)=M_{\text{cold}}(t)-M_{\text{cold}}(0)$, for the AMR simulations relative to their static grid counterparts. We use this quantity rather than the ratio of the total cold gas masses in order to make the rather small differences discernible. The ratio of the total cold gas masses will always be very close to unity due to the initial cold gas in the cloud dominating the mass. We compare AMR simulations to static grid simulations using both divergence cleaning and CT. As can be seen, the differences are typically within about 10\%. For the HD cases, AMR leads to slightly higher cold gas masses than their corresponding static grid cases at any time. In the low resolution cases, there is virtually no difference between AMR and static grid runs until the last several Myr. For the strong magnetic field case, the difference in $\Delta M_{\text{cold}}(t)$ between the AMR and static runs is not significant either but fluctuates more. In most cases, the difference is greater at standard resolution than at low resolution. In general, the difference between AMR and static grid runs at standard resolution is smaller whenever an ambient magnetic field is present, even though the differences are quite minor in the pure HD case as well.

Comparing specifically the divergence cleaning and CT cases, although the evolution of their relative cold gas masses is not congruent, the difference is less than 10\% at all times. The magnetic field shows more small scale structure when CT is used rather than divergence cleaning. Specifically, two vortices form at the end of the wake in both cases but are more prominent in the CT case. Also, some magnetic filaments form along the $y$ direction in the cloud's wake in the CT case but are absent when divergence cleaning is used. The density distribution of the wake is also different as a result. While in both the AMR and static mesh simulations with divergence cleaning the wake is shaped as a single tail, the end part of the tail splits into two when CT is used. Comparing the field in simulation 2HS to 2HS-s (i.e. where divergence cleaning is used in both cases) shows that the AMR run again contains less small scale structure. Thus, both divergence cleaning and AMR appears to smooth out the field to some degree. However, as previously mentioned, this smoothing and the corresponding changes in wake morphology do not lead to any significant differences in the overall amount of cold gas condensation. This is because the essential aspects of the magnetic field, in particular the amplification of the field at the cloud's leading edge, are still preserved. Thus, for our purposes, divergence cleaning leads to an appropriate approximation of the solenoidal field.

%--------------------------------------------------------------------------------------------------------------------------------------------------------------------------------
%--------------------------------------------------------------------------------------------------------------------------------------------------------------------------------

\subsection{Effect of dimensionality}
\label{sect:2Dvs3D}

The results presented by \citetalias{marinacci10} and subsequent work \citep{marinacci11,marasco13,armillotta16} where the phenomenon of GF assisted condensation was explored were based mainly on two-dimensional (2D) simulations. \citetalias{marinacci10} in particular argued that this approximation led to an overestimate of gas mixing and, as a consequence, of the amount of gas condensation. Their key argument is that the power spectrum of the turbulence -- essentially a power law -- is shallower in 2D compared to the full 3D problem. However, \cite{armillotta16} arrived at the opposite conclusion using a full 3D simulation --- that is, they found that the gas condensation is {\em higher} in 3D compared to 2D, although it must be noted that the difference in cold gas mass between their experiments was relatively modest.

We argue that the latter result is closer to reality. We speculate that the reason is that in the 2D case, a `spherical' cloud effectively corresponds to the cross section of an infinitely long cylinder, and its surface area to a one-dimensional `ring,' thus having a substantially smaller area normal to the wind compared to a true 3D sphere (or any other 3D object for that matter). This reduces the chances for instabilities to develop, which in turn reduces the mass of gas that is ablated and that eventually mixes with the ambient medium, thus leading to a lesser amount of condensation.

%--------------------------------------------------------------------------------------------------------------------------------------------------------------------------------
\begin{figure}[h]
\begin{center}
\includegraphics[width=0.49\textwidth]{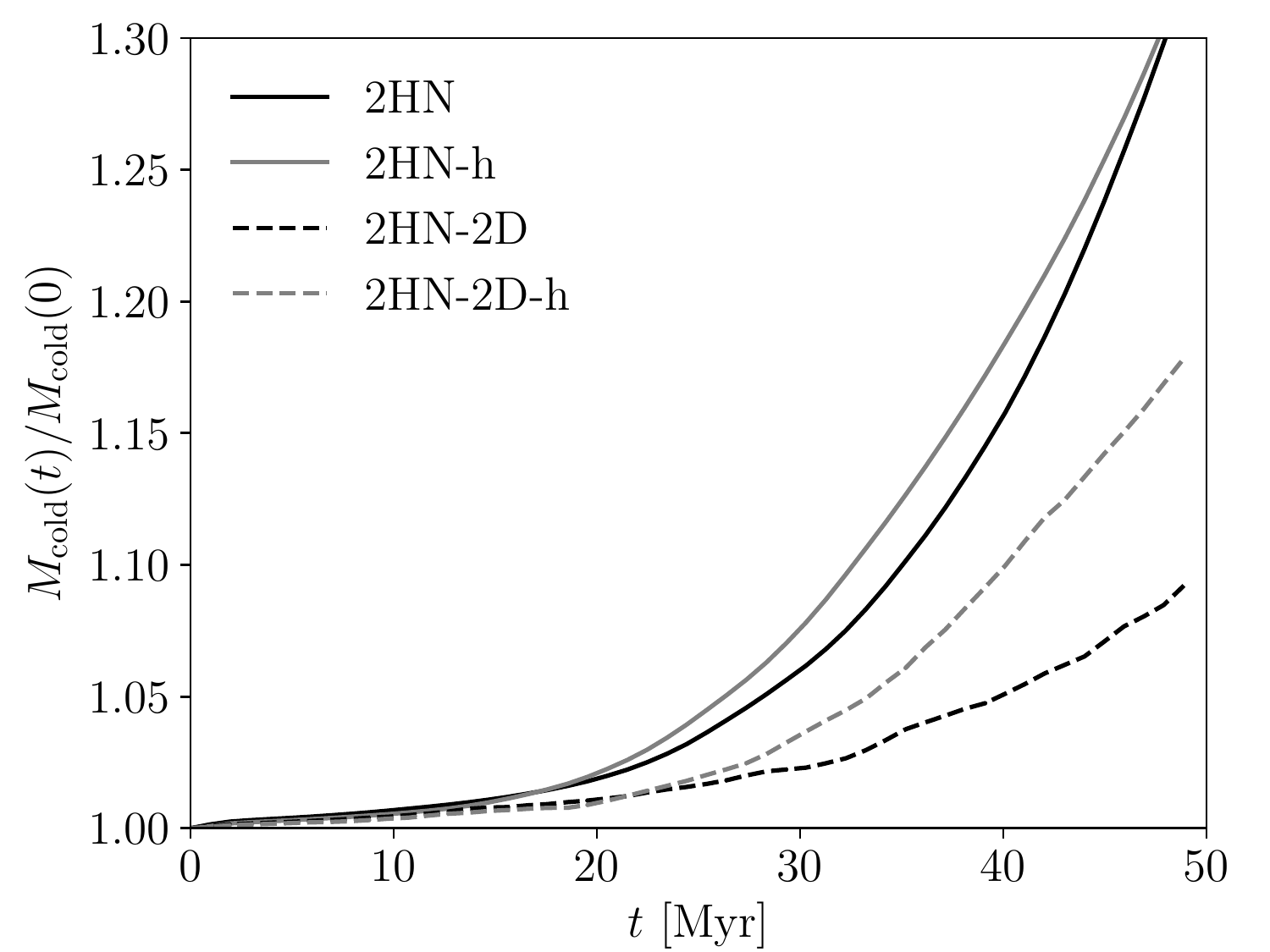}
\end{center}
\caption{Gas condensation for the 3D simulations 2HN (solid black) and 2HN-h (solid gray) and their corresponding 2D versions 2HN-2D (dashed black) and 2HN-2D-h (dashed gray), respectively.}
\label{fig:2dvs3d}
\end{figure}
%--------------------------------------------------------------------------------------------------------------------------------------------------------------------------------

In order to asses the impact of dimensionality on the resulting amount of condensed gas quantitatively, we compare the evolution of the mass of condensed gas over the course of simulations 2HN and its high-resolution counterpart 2HN-h (see Section~\ref{sec:resolution} for a discussion of resolution in the 3D simulations) with identical simulations 2HN-2D and 2HN-2D-h, respectively, which are run in 2D. The result of this exercise is shown in Figure~\ref{fig:2dvs3d}. Our results are consistent with \cite{armillotta16} in that gas condensation is generally higher in 3D compared to 2D at all times. It is worth emphasizing that the difference is negligible at earlier times, though, but quite significant at later times. The evolution of the cold gas mass in 2HN-2D-h follows that of its lower resolution counterpart until about $t=30$ Myr, after which the cold gas mass grows significantly more rapidly at the higher resolution. This is somewhat different from the three-dimensional case, where the increased resolution only leads to slightly more condensation. As previously mentioned, simulation 2HN-h becomes affected by numerical issues at late times. In any case, it is clear that ignoring the third dimension leads to a significantly lower condensation rate.

Note that we do not compare MHD simulations in 2D and 3D. These would obviously be quite different from one another because the presence of a uniform (or otherwise regular) magnetic field breaks the symmetry of the system, as has long been recognized \citep{gregori99}, which leads to a substantially different evolution along different axes \citep[see  Section~\ref{sec:supp} and ][]{gronnow17}.

\section{Discussion}
\label{sec:discussion}

\subsection{Magnetic suppression of instabilities}
\label{sec:instabilities}
The suppression of the KH instability by magnetic fields can be derived analytically under simplifying assumptions \citep{chandrasekhar61}. The typical timescale for the simplest case of linear perturbations in an {\em incompressible}, uniformly magnetized flow along a parallel interface to grow is
\begin{equation}
\label{eq:khtimescale}
t_{\text{KH}} = \left(\frac{\rho_1\rho_2}{(\rho_1+\rho_2)^2}(\mathbf{k} \cdot \mathbf{v})^2 - \frac{2(\mathbf{B}\cdot \mathbf{k})^2}{\rho_1+\rho_2}\right)^{-1/2}
\end{equation}
where $\mathbf{v}$ is the shear velocity across the interface, $\rho_1$ and $\rho_2$ are the densities on either side of the interface, $\mathbf{B}$ is the magnetic field, and $\mathbf{k}$ is the wave vector of the perturbation. As emphasized previously, the expression provided may not be valid for compressible fluids such as the halo-cloud system in our simulations \citep{karimi16}. Still, it provides some insight into the effect of magnetic fields on the onset of the instability and the results of our experiments. Quite generally, the greater the magnetic field along the direction of the dominant mode, the greater the characteristic timescale needed for this mode to grow. Due to the draping (or bending) of the magnetic field lines embedded in the fluid, the term $\mathbf{B}\cdot \mathbf{k}$ will not vanish, in general, along the contact surface \citep{banda-barragan16}. This is true regardless of the initial field orientation (see Section \ref{sec:oblique}). Thus, the presence of a magnetic field will generally dampen the growth of KH modes, as we have argued before.

In principle, the cloud should be stable against {\em linear} KH instability for $\mathcal{M}_A \lesssim 2$ \citep{ryu00}, as is the case for our strong field simulations at $v_{\text{wind}}=75 \kms$. However, it might still operate on smaller scales due to local fluctuations in the Alfv\'{e}n speed. More importantly, KH is not the only instability affecting the clouds. RT instability caused by strong density variations is important in the final break-up of clouds \citep{banda-barragan16}. The RT instability should be subdominant to KH instability, though, during all of our non-magnetic simulations, except for the high-velocity case (see Section \ref{sect:highvel}). However, when there is a magnetic field, it might become the dominant source of gas ablation and consequent gas mixing because the uniform magnetic field does not suppress, and can even enhance, RT instability modes in the $z$ direction \citep{gregori99,gronnow17}. The morphology of the cloud's wake in simulation 2HS clearly shows this effect (see Figure \ref{fig:rhocoldprojs}). The tail is thin and highly collimated in the $xy$-plane but much more spread out along the $z$-axis. The slightly enhanced condensation in the strong transverse field case compared to the parallel field case could also be a consequence of an enhanced RT instability. In addition, the magnetic field is the source of the tearing mode (TM) instability \citep{furth63}. This is caused by reconnection and so should technically not be present in ideal MHD, but it still occurs to some extent in grid-based MHD simulations because of numerical resistivity \citep{rembiasz17}.

\subsection{Magnetic field draping}
\label{sec:draping}
The qualitative differences between our weak and strong field simulations can be understood through the magnetic field `draping' phenomenon. The effect of the cloud's motion on the magnetic field depends on its Alfv\'{e}nic Mach number. When $\mathcal{M}_A \gg 1$ the magnetic field drapes around the cloud in a thin, highly amplified layer (see \cite{dursi08} for a discussion of this phenomenon). This effect is greatest when the field ahead of the cloud is at a right angle to the cloud's motion. In addition to having an enhanced field strength, the draping layer generally follows the direction of the flow, thus maximizing the $\mathbf{B} \cdot \mathbf{k}$ term in the KH instability timescale (eq. \ref{eq:khtimescale}). Our weak field simulations are in this highly super-Alfv\'{e}nic regime with $\mathcal{M}_A=8.2$ while our strong field simulations are sub-Alfv\'{e}nic with $\mathcal{M}_A=0.8$ because $v_A \propto \vert\mathbf{B}\vert$. The field draping in the weak field case, and its absence in the strong field case, can be clearly seen in Figure \ref{fig:bfieldslices}, where the strong field is amplified mainly at the cloud's leading edge and is bent rather than draped.

Several of our results can be explained through the draping effect and its dependence on $\mathcal{M}_A$:\\

(i) In simulation 2HW, the condensation of cold gas is suppressed more strongly than in simulation 2HS even though the magnetic field is initially weaker. This is caused by the high field amplification around the cloud and its wake as a result of draping. The latter is absent in simulation 2HS except at the cloud's leading edge. Of course, if the initial field becomes progressively weaker, at some point the amplification can no longer compensate for the overall weak field, and the cloud's evolution asymptotes to the pure HD case.

(ii) In our weak field simulations, the strength of the magnetic field around the cloud, and in turn the suppression of gas stripping and cold gas condensation, increases monotonically as the angle of the field relative the $y$-axis is increased, because the draping increases correspondingly. Since the strong field is not draped at any angle, but rather only bent, the orientation of the field has a different effect (see Section \ref{sec:oblique}).

(iii) In the transonic simulations, the condensation is lower in the strong field case compared to the corresponding subsonic simulation up until $t \approx 25$ Myr. However, the condensation is higher at all times in the transonic, weak field case compared to the subsonic run. Also, in the transonic, strong field case, the wake has a morphology that more closely resembles that of the weak field case than that of the subsonic, strong field case. This is because at the higher speed the cloud is super-Alfv\'{e}nic in both the weak and strong field cases. Thus the field drapes around the cloud in the high-velocity, strong field case, leading to significant field amplification, which in turn effectively largely inhibits the condensation of gas along the cloud's wake. In contrast, when a weak ambient field is present, the draping and amplification already occurs (and saturates) at low velocity. Therefore the increased velocity does not lead to significantly higher field strengths around the cloud and wake. As a result, at higher velocity the velocity dependent term in eq. \ref{eq:khtimescale} grows while the magnetic term does not. This causes the KH timescale to become shorter, in addition to a stronger shock and additional RT instability. This explains why for the weak field, the higher velocity only increases the condensation.

%--------------------------------------------------------------------------------------------------------------------------------------------------------------------------------
\begin{figure*}[t]
\begin{center}
\includegraphics[width=0.7\textwidth]{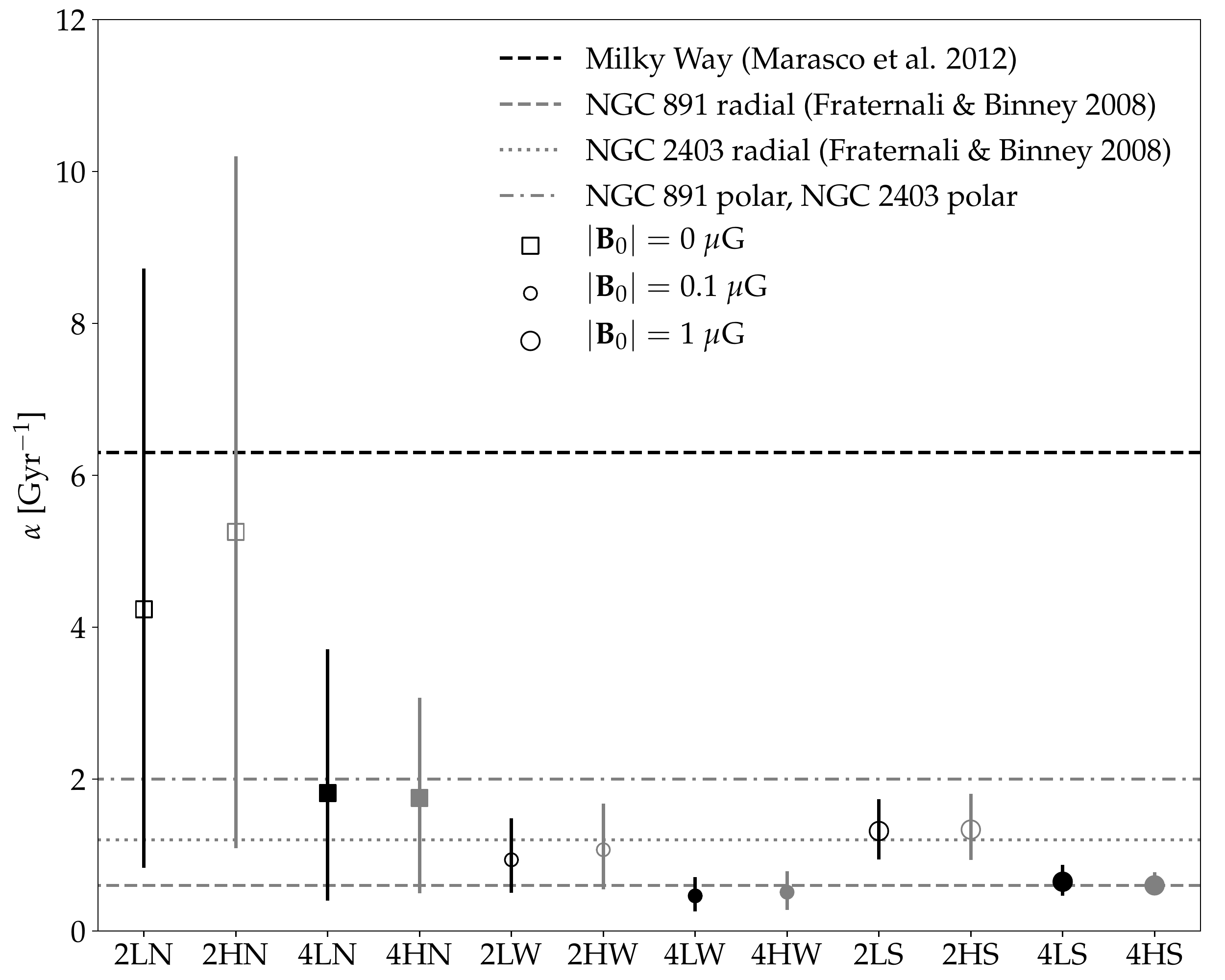}
\end{center}
\caption{Specific condensation rate $\alpha$ for our standard simulations compared to constraints from statistical Galactic fountain galaxy accretion models matched to observations of \textsc{Hi} gas kinematics. The horizontal axis labels are the names of the simulations, so it is not a continuous quantity and the ordering of the sequence is not important. $\alpha$ is defined from the assumption in these models that $M_{\textrm{cold}(t) \sim \exp{(\alpha t)}}$. The gray dashed and gray dotted horizontal lines represent lower bounds on $\alpha$ for the Milky Way like galaxies NGC 891 and NGC 2403, respectively, from \cite{fraternali08}. The gray dotted-dashed line represents the upper bounds of NGC 891 and NGC 2403, which are similar. The black dashed line is the estimate of $\alpha$ for the Milky Way from \cite{marasco12}. Open (filled) symbols represent low (high) density simulations; black (gray) symbols represent low (high) metallicity simulations. The symbols represent $\alpha$ fitted to the cold gas mass from $t=0$ to $t\approx 50$ Myr, while the bars span the range from fitting to only the first 25 Myr to only the last 25 Myr of the simulations.}
\label{fig:alpha}
\end{figure*}
%--------------------------------------------------------------------------------------------------------------------------------------------------------------------------------

\subsection{Random magnetic field components}
In this paper, we only consider uniform ordered magnetic fields. However, the magnetic field in the Galactic halo likely has a significant random component, caused by local turbulence, in addition to the large-scale ordered field \citep{jansson12b}. In the cloud-wind simulations of \cite{mccourt15} and \cite{banda-barragan18} the clouds had internally (partially) random fields while the field in the surrounding medium was uniform. They found that these clouds experience less severe ablation compared to the non-magnetic case, indicating that the suppression of the KH instability is still significant in this case. However, in general, in addition to the cloud having its own partially random internal magnetic field, the halo field will have a random component as well. For our strong magnetic field case, where draping and amplification is not important, a less ordered field should still lead to a non-negligible suppression of the KH instability around the cloud because the component of the field pointing along the cloud-halo interface will still generally be significant. For our weak magnetic field case, the direction of the ordered component will still be important but the random component should ensure that some amplification of the field occurs around the cloud, even if the ordered component is parallel to the cloud's motion.

\subsection{Comparison to observations}
\label{sec:comparisonobs}
We compare the condensation in our simulations to the observationally derived estimates from the statistical models of Galactic fountain accretion of \cite{fraternali08} and \cite{marasco12}. In these models, clouds represented as single particles are ejected from the disk on ballistic trajectories at a rate proportional to the local SF rate. They assume that condensation causes the mass of these clouds to increase at a rate $\dot{M}=\alpha M$ while traveling through the circumgalactic medium. This leads to an exponential growth in cold gas mass for each cloud $M_{\textrm{cold}}(t) \propto \exp{(\alpha t)}$. \cite{fraternali08} estimated the (constant) specific accretion rate $\alpha$ (in units of Gyr$^{-1}$) for the two Milky Way like galaxies NCG 891 and NGC 2403 constrained by observations of the kinematics of extra-planar \textsc{Hi} gas in these systems. They found best fitting values of order unity, and which are consistent with the average SF rates measured in these galaxies. \cite{marasco12} applied an extension of this model to the Milky Way and found a best fitting value of $\alpha=6.3$. In Figure \ref{fig:alpha} we show values of $\alpha$ estimated for our basic set of simulations and the observationally derived estimates. We estimate $\alpha$ in our simulations by fitting an exponential of the form $M_{\textrm{cold}}(t)=M_0 \exp{(\alpha t)}$ to the cold gas mass evolution. For the observational constraints from \cite{fraternali08}, two extreme values are given for each galaxy corresponding to predominately radial and predominately polar flow, giving a range for $\alpha$. In both cases, the upper bound (polar flow) is at $\alpha=2.0$ Gyr$^{-1}$. We find that the evolution of the cold gas mass in our simulations is generally much better fitted by a quadratic rather than by an exponential function. This was also the case for the 3D simulation of \citet[see their Figure 7]{armillotta16}. As a consequence, $\alpha$, being based on an exponential fit, is not constant but rather changes based on which part of the cold gas mass evolution is included in the fit. Fitting to the late stages of the evolution generally leads to an increase in $\alpha$ compared to the early stages. Therefore, in addition to showing $\alpha$ from fitting an exponential to the evolution from $t=0$ to $t\approx 50$ Myr, we also derived early and late estimates based on only fitting $\alpha$ to the first and last 25 Myr, respectively. As can be seen, the ranges for all our simulations overlap with the observational values. As perhaps expected based on our previous discussion, $\alpha$ is significantly higher when there is no magnetic field. While they are still mostly consistent with observations, at late times, the low-density clouds are accreting gas at a somewhat too high rate. This is only a lower bound, because typical fountain cloud travel times are $\sim 80$ Myr \citep{marasco13}. Indeed, for simulation 2HN, which we do run until $t\approx 80$ Myr, we find that $\alpha=13$ when fitting to the last 40 Myr and $\alpha=10$ when fitting to the overall $t=0 - 80$ Myr evolution. The two corresponding MHD simulations that we have also run until $t\approx 80$ Myr have late stage values of $\alpha$ less than 4, still well within the observational constraints.

In brief, full 3D, pure HD simulations significantly {\em overpredict} the value of $\alpha$ compared to its value as inferred from observations. The presence of an ambient magnetic field brings down the specific accretion rate back to values consistent with the data, thus alleviating this issue.

\subsection{Comparison to previous studies and implications for Galactic gas accretion}
\label{sec:comparison}

As shown in Section~\ref{sec:results}, the presence of an ambient, initially uniform magnetic field generally has a significant negative impact on the overall condensation of cold gas along the wake of gas clouds. We have shown that this effect is, qualitatively, robust to reasonable variation of the relevant physical and numerical parameters.

Our simulation 2HN is comparable to the model dubbed `11' in \cite{armillotta16}, except for the lower metallicity of the halo in their case, and the inclusion of thermal conduction. They found that thermal conduction also inhibits the condensation of gas, although its effect is not as significant as the effect of a magnetic field, based on our results. Using the same definition of cold gas of $T<2\times 10^4$ K as they did, at the end of their simulation at $t=60$ Myr, we find $M_{\textrm{cold}}(t=60 $Myr$)/M_{\textrm{cold}}(0)=1.6$ in simulation 2HN which is significantly higher than their result of $\approx 1.3$, which is expected given their lower halo metallicity of 10\% solar and the suppression caused by thermal condition. In contrast, $M_{\textrm{cold}}(t=60 $Myr$)/M_{\textrm{cold}}(0)=1.15$, for both the corresponding weak and strong field simulations (2HW and 2HS).

While we have chosen the initial conditions of our simulations to be broadly consistent with the ones used by \citetalias{marinacci10}, it is not possible to exactly replicate theirs. For example, the cloud : halo mass density contrast in all our simulations is roughly a factor of two higher than in \citetalias{marinacci10}, even though we use similar number densities. This is due to their assumption of a constant mean molecular weight, while we take its temperature dependence into account. Overall, the mass of condensed gas in our 3D MHD simulations is comparable to that of the 2D, pure HD simulations of \citetalias{marinacci10} and \cite{marinacci11}. But as our simulations are not converged with respect to spatial resolution, and the efficiency of condensation appears to decrease with resolution in the MHD case, the amount of condensed gas resulting from our experiments should be taken as a strict upper limit.
As we show in Section \ref{sec:comparisonobs} the amount of condensation is too high at late times in some of our simulations without magnetic fields compared to observational constraints on Galactic fountain accretion rates. We also show that using full 3D simulations increases the amount of condensation significantly compared to 2D, because the 2D geometry artificially suppresses the KH instability (see Section \ref{sect:2Dvs3D}).

We used a smooth initial density profile for the cloud, which aids the numerical stability early in the simulations. A smoother profile should suppress the KH instability compared to a sharp one, as was used in \citetalias{marinacci10} and \cite{armillotta16}. However, our initial profile is quite steep and quickly becomes further steepened, making the cloud-halo transition region only a couple of cells wide. This steepening is caused by the low-density outer parts of the cloud being swept away and the more inner parts being compressed by the wind. Therefore any differences between our results and these previous studies caused by the different density profiles will be negligible compared to the other differences in the initial conditions previously mentioned.

Our results, taken collectively, together with the presence of a magnetic field in the Galactic corona and the disc-halo interface \citep[qv.][]{beck16}, show that the magnetic field affects condensation so severely that it is critical to take it into account in numerical and theoretical studies of any type of cloud-driven accretion.

\citetalias{marinacci10} find that a higher overall gas metallicity substantially increases the amount of gas condensation. We find that uniformly increasing the gas metallicity leads to more condensation as well, but the effect is relatively small for the metallicity range of $Z\approx 0.3 Z_\odot$ to solar that we examine. The effect of changing only the cloud's metallicity is more complicated (see Section \ref{sec:metal}).

Based on our result that the presence of an ambient magnetic field leads to less condensation compared to a pure HD case, even for relatively weak magnetic fields ($\vert \mathbf{B}\vert=0.1 \mu$G, $\beta=600$) and high velocities ($v=200 \kms$), we speculate that a similar effect would be present in the case of Galactic HVCs. The phenomenon of cold gas condensation associated with HVCs has been studied with the help of numerical experiments by \citetalias{gritton17}. They simulated massive, unmagnetized HVCs moving slightly subsonically to slightly supersonically through the halo for typically $\sim 100$ Myr. They found that these clouds can accrete significant amounts of cold gas, in the most extreme case nearly doubling the initial cold gas mass after about 160 Myr. In all their simulations, the cold gas mass increases monotonically throughout (barring mass loss from gas flowing out of their simulation domain at late times). Their initial conditions were quite different from ours, and the massive HVCs that they examined could still condense a significant amount of gas over the longer timescales associated with the journey of HVCs through the halo. The amount of mass actually accreted through this condensation will depend on how far the HVC travels before either dispersing completely or merging with the disk.

We note that \citetalias{gritton17} do not address the question of how the cold gas that typically condenses 10s of kpc above the disk in the wake of the HVCs may survive the journey to the disk. As the density of the surrounding halo gas increases along the trajectory, it is expected that some fraction of this gas will disperse, fully mixing with the hot material, rather than being accreted onto the disk \citep[e.g.,][]{tepper-garcia15}. It is therefore unclear for now whether the condensed gas will actually rain down onto the disk. More comprehensive simulations that include density and magnetic field gradients along a cloud's orbit through the corona, as well as the Galaxy's gravitational potential, are required to address this properly. A related issue is how the magnetic field may affect the survival of cold gas accreting onto the disk. Generally, the relatively strong field at the disk-halo interface may stall the inflow of gas onto the disk because infalling gas has to reach a critical mass to overcome the magnetic tension \citep{birnboim09}.

The effect of uniform magnetic fields in super-Alfv\'{e}nic cloud-wind and cloud-shock interactions has been examined in many previous numerical studies, going back to the seminal works of \cite{maclow94} and \cite{jones96}. While these early studies were restricted to 2D simulation domains, they also observed the stabilizing effect of the field on the cloud. \cite{jones96} argued that the efficiency of draping depends on the sonic Mach number. However, we find, in agreement with \cite{dursi08}, that it depends instead primarily on the Alfv\'{e}nic Mach number. \cite{banda-barragan16} tested the effect of different field orientations on the evolution of wind-swept clouds in the strongly super-Alfv\'{e}nic regime, using full 3D MHD simulations. Their results agree with ours whenever their and our explored parameter spaces overlap.

We find that the condensed gas mass is not converged at our standard resolution of $\mathcal{R}_{64}$. This is consistent with previous findings for cloud-wind simulations with radiative cooling. In their study of clouds embedded in high-velocity supernova outflows, \cite{cooper09} found no convergence in the amount of cloudlets formed in the wake of a cloud for resolutions up to $\mathcal{R}_{50}$. Examining a similar physical setup, \cite{scannapieco15} found that $\mathcal{R}_{64}$ was sufficient to resolve the mass retained in the cloud and the average velocity of cloud material. However, the mass fraction of the multiphase gas was not converged, suggesting that they were not completely resolving the KH instability at their adopted resolution. In a study of a cloud-wind interaction of clouds with an initially fractal density distribution, \cite{schneider17} found moderate differences between $\mathcal{R}_{64}$ and $\mathcal{R}_{128}$ for the mass retained by the cloud.

\citetalias{marinacci10} found that the amount of condensation was highly dependent on resolution and did not converge at their maximum resolution of $\mathcal{R}_{67}$. However, these simulations were in only two dimensions, as discussed in Section~\ref{sect:2Dvs3D}. \citetalias{gritton17} found that $\mathcal{R}_{32}$ was sufficient to resolve the mass of condensed gas for their 3D AMR simulations of HVCs.

\section{Summary}
\label{sec:summary}
Analytic models of accretion based on the galactic fountain process \citep[e.g.,][]{fraternali08,marasco12} have been successful in fitting the kinematic properties of extra-planar \textsc{Hi} gas. In these models, it is assumed that clouds  moving through the lower Galactic corona increase their mass at an exponential rate. Previous studies based mainly on 2D hydrodynamic simulations of cloud-wind interactions (e.g. \citetalias{marinacci10}; \citealt{marinacci11}; \citealt{armillotta16}) showed that the condensation of gas in an amount required by observations can be triggered along a fountain cloud's wake as a result of gas ablation and mixing.

In this work, we have revisited these claims, running extensive simulations both in 2D and 3D, and both neglecting and including the presence of an ambient magnetic field. We find that 2D simulations artificially suppresses the KH instability at the cloud-halo interface. This in turn reduces the ablation of clouds, thus decreasing the amount of gas that mixes with the ambient medium, which is the process that ultimately drives the condensation of gas. Thus, 2D simulations systematically underestimate the amount of condensed gas. At the other extreme, full 3D, non-magnetic simulations generally result in significant amounts of cold condensed gas, too high to be consistent with constraints on Galactic fountain accretion rates inferred from observations \citep{fraternali08,marasco12}. We find that the presence of an ambient magnetic field --- as has been observed around the Milky Way and other spiral galaxies \citep{beck16} --- strongly suppresses the condensation of gas along a cloud's wake. The magnetic field effectively inhibits the onset (i.e. dampens the growth of) hydrodynamic (KH) instabilities. This leads to a reduction in the amount of condensation for the same reason as the reduction seen in 2D simulations. However, we claim that unlike the 2D case, this is a physical, rather than numerical, effect. Thus, the magnetic field alleviates the problem of condensation being too efficient in our 3D simulations, bringing the accretion rate down to be in good agreement with observational constraints. Significant suppression occurs even for quite weak initial fields ($\beta\approx 600$). The key mechanism behind this effect is the amplification of the field ahead of and around the cloud. We show that the field amplification is directly related to the phenomenon of `draping,' which in turn depends on the relative orientation of the field and the Alfv\'{e}nic Mach number of the cloud.

This suppression of gas condensation by the presence of a magnetic field is found to be generic and universal, as it is qualitatively robust to reasonable variations of the values of physical parameters such as the metallicity, density, and velocity of the cloud, or the strength and orientation of the magnetic field, or values of numerical parameters such as the limiting spatial resolution and the method used to minimize magnetic field divergence.

Our results highlight the importance of magnetic fields in processes that rely on gas mixing, such as galactic fountain driven accretion. We stress that magnetic fields cannot be ignored if we are to arrive at a full understanding of this process.

%--------------------------------------------------------------------------------------------------------------------------------------------------------------------------------
\section*{Acknowledgements}
We thank the referee for valuable insights and useful suggestions that improved our paper.
A.G. and T.T.G. acknowledge financial support from the Australian Research Council (ARC) through an Australian Laureate Fellowship awarded to J.B.H. We acknowledge the facilities and the scientific and technical assistance of the Sydney Informatics Hub at the University of Sydney, and in particular, access to the high-performance computing facility Artemis and additional resources on the National Computational Infrastructure (NCI) through the University of Sydney's Grand Challenge Program `Astrophysics Grand Challenge: From Large to Small' (CIs G. F. Lewis and J. Bland-Hawthorn). We also acknowledge additional access to NCI facilities through the Astronomy Supercomputer Time Allocation Committee (ASTAC) scheme managed by Astronomy Australia Limited and supported by the Australian Government.

\software{PLUTO \citep[version 4.1 of the code last described by][]{mignone07,mignone12}; VisIt \citep{HPV:VisIt}}.

%--------------------------------------------------------------------------------------------------------------------------------------------------------------------------------
%\bibliographystyle{apj}
\bibliography{AG_condensation}
%--------------------------------------------------------------------------------------------------------------------------------------------------------------------------------

%--------------------------------------------------------------------------------------------------------------------------------------------------------------------------------
\end{document}